\documentclass[
 prd,aps,
 amsmath,amssymb
]{revtex4-2}

\usepackage{amsmath}
\usepackage{amsthm}
\usepackage{needspace}
\usepackage{graphicx}
\usepackage{dcolumn}
\usepackage{bm}
\usepackage{hyperref}
\usepackage{bigints}
\usepackage{setspace}
\usepackage{color}
\theoremstyle{plain}

\newtheorem{assumption}{Assumption}
\newcommand{\scri}{\mathcal{I}}
\newcommand{\Arad}{\mathcal{A}_{\mathrm{rad}}}
\newcommand{\Agrav}{\mathcal{A}_{\mathrm{grav}}}

\newcommand{\QB}{Q_{B}}
\newcommand{\Hfp}{\mathcal{H}_{f^{+}}}
\newcommand{\Hfm}{\mathcal{H}_{f^{-}}}
\newcommand{\Htot}{H_{\mathrm{tot}}}
\newcommand{\QBL}{\mathcal{Q}^{L}_{B}}
\newcommand{\Agravtwo}{\mathcal{A}^{(2)}_{\mathrm{grav}}}
\begin{document}

\title{Asymptotic Algebras and Holography of Information in CGHS Model}

\author{Waheed A. Dar}
\email{waheed.dar1729@gmail.com}
\affiliation{Department of Physics, National Institute of Technology Srinagar,\\
Kashmir-190006, India}

\author{Nirmalya Kajuri}
\email{nirmalya@iitmandi.ac.in}
\affiliation{School of Physical Sciences, Indian Institute of Technology Mandi,\\
Himachal Pradesh, India}

\author{Rinkesh Panigrahi}
\email{d23189@students.iitmandi.ac.in}
\affiliation{School of Physical Sciences, Indian Institute of Technology Mandi,\\
Himachal Pradesh, India}

\begin{abstract}
Holography of Information (HoI) and the island formula are two recent approaches to the black hole information loss paradox that yield different Page curves. The difference between the Page curves has been argued to be rooted in the algebra that each approach focuses on. In this paper, we study the candidate algebras for the CGHS model. For the gravitational completion constructed here and under assumptions analogous to those made in previous HoI literature, we establish HoI for the CGHS model coupled to a massless scalar by constructing the radiative phase space and asymptotic algebras at future null infinity. We prove that the quantum state of the right-moving modes can be recovered from an arbitrarily small neighbourhood of the past endpoint of the right future null infinity. We then argue that island reconstruction is incompatible with exact commutativity of the left and right radiative algebras. We discuss the possible distinction between the HoI and island approaches for the CGHS model. 
\end{abstract}
\maketitle

\section{Introduction}

The black hole information loss problem remains one of the central puzzles at the interface of quantum mechanics and gravity. Hawking's original calculation showed that a black hole formed from pure-state collapsing matter appears to radiate thermally and evaporate to a mixed state, in an apparent violation of unitarity \cite{Hawking:1976ra}. It has long been expected that the solution to the paradox will show that the Page curve of radiation entropy rises and then falls back to zero \cite{Page:1993df,Page:1993wv}.

Two recent approaches have shed new light on this question. But they appear to give different answers with respect to the Page curve.

Holography of Information (HoI) \cite{Laddha:2020kvp,Chowdhury:2020hse,Chowdhury:2021nxw,Laddha:2022nmj,Chakraborty:2023los,Jensen:2024dnl,Gaddam:2024mqm}, which developed earlier ideas from \cite{Banerjee:2016mhh,Raju:2018zpn,Raju:2019qjq}, takes as starting point the algebra of asymptotic gravitational observables. The central result of this approach is that the quantum information  of massless excitations is stored in an arbitrarily small neighbourhood of the past boundary of future null infinity. Thus the graph of von Neumann entropy of such states remains constant with the affine parameter on future null infinity. Put differently, it shows that the Page curve should be flat.

In contrast, the island formula, derived from the quantum extremal surface prescription \cite{Almheiri:2019psf,Penington:2019npb}, produces a Page curve that rises and then falls. The key finding here was that the fall is driven by an ``island" region extending into the black hole interior \cite{Almheiri:2019hni,Penington:2019kki, Almheiri:2019qdq, Almheiri:2020cfm,Almheiri:2019yqk, Almheiri:2019psy}.  

The contrast between the results obtained by the two approaches has been discussed both from the island \cite{Antonini:2025sur} and the HoI \cite{Geng:2026asi} perspectives. 
Both these papers suggest that the difference between the results stems from the difference in the asymptotic algebras that the two approaches focus on. The island formula computes the entropy of the algebra of radiative observables, while HoI computes the entropy of its gravitationally completed extension that includes the Bondi mass at every cut. Where disagreement persists is the question: which of these algebras is physically appropriate to describe the escape of information in the context of the black hole information paradox? 

A useful arena for examining this question is provided by the CGHS model \cite{Callan:1992rs} coupled to a massless scalar. The CGHS model is asymptotically flat and has long served as a useful arena for studying of black hole information loss \cite{Giddings:1992ff,deAlwis:1992hv,Giddings:1992ae,Bilal:1992kv,Strominger:1994tn,Gegenberg:1994pv,Varadarajan:1997qz,Varadarajan:2008zz,Ashtekar:2008jd,Ashtekar:2010hx,Ashtekar:2010qz,Almheiri:2013wka}. The advantage for our purposes is the simplification in asymptotic algebras.

Our aim in this paper is twofold. First, we establish CGHS analogues of
the HoI results that Laddha et al. \cite{Laddha:2020kvp} established for asymptotically flat 4-dimensional gravity. By this we mean completeness
of the specified gravitational algebra near the past endpoint or
endpoints, the inclusion of later-cut information in earlier-cut
algebras under the stronger finite-cut assumptions, and the associated
entropy statements. Our second aim is to compare the asymptotic algebras corresponding to the island and HoI approaches in this context. We will reexamine the conclusion that the two approaches is restricted to the difference of algebras.

A peculiar feature of CGHS model is that the massless matter field divides into left- and right-moving modes, which approach the left and right future null boundaries respectively. We work in the setting where the left-moving mode collapses to form the black hole, while the right-moving mode plays the role of Hawking radiation. As was observed in \cite{Almheiri:2013wka}, for the CGHS model the information loss problem divides into two separate problems. One is the retrieval of the information that went inside the black hole, carried by the  left-moving modes that approach at $\mathcal{I^+_L}$. The other is the retrieval of the state of Hawking radiation, carried by the right-moving modes which approach $\mathcal{I^+_R}$.

In this paper, we will consider two cases. One where HoI analysis is restricted to only the right boundary. That is, restricting to only the second problem. The other case is where both boundaries are considered. 

To establish HoI, we first construct the radiative phase space for the CGHS model at the right null infinity and obtain the boundary Hamiltonian. Notably, the radiative data at the null boundary is that of the matter field
alone. This simplifies the identification of the radiative algebra. However, unlike the four-dimensional case, there is no gravitational news and thus no unique choice of the asymptotic gravitational algebra. We construct the gravitational algebra by appending charges that we construct from the asymptotic coefficients of the dilaton. This choice is motivated by the 4-dimensional case, and not claimed to be unique or maximal. It is found to reproduce the HoI results.

By choosing this algebra and making assumptions similar to the ones made in \cite{Laddha:2020kvp} for 4-dimensional asymptotically flat spacetimes (the crucial one being that the black hole evaporates, leaving no horizons or singularities), we establish results analogous to those of \cite{Laddha:2020kvp}. Specifically, quantum information of the state of Hawking radiation on the right future null boundary can be retrieved from an arbitrarily small neighbourhood of the past of the right future null boundary. Thus a flat Page curve will be obtained. This holds true both for the one-boundary restriction and the two-boundary case. 

We further argue that this result will continue to hold even when the horizon is present. Then the state on the right future null infinity is mixed. Nevertheless, the information of this state can be retrieved from an arbitrarily small neighbourhood in the far past.

The one-boundary setting yields interesting no-go result concerning islands. As emphasized in \cite{Antonini:2025sur}, semiclassical operators in an island can be reconstructed by operators acting on the Hawking radiation. If the island contains left-moving modes, a noncommuting pair of left-moving operators that are also represented in the radiative algebra of observables at the left boundary $\mathcal A_{\rm rad}(\mathcal I_L^+)$ must therefore be reconstructible in  the radiative algebra of observables at the right boundary $\mathcal A_{\rm rad}(\mathcal I_R^+)$. 

Such a reconstruction is incompatible with exact commutativity of $\mathcal A_{\rm rad}(\mathcal I_R^+)$ and $\mathcal A_{\rm rad}(\mathcal I_L^+)$. A right-radiation reconstruction would commute with the left radiative algebra, whereas the corresponding left-moving operators do not commute among themselves.  

There are then two possibilities. One is that the island formula does not apply to the CGHS model. The other is that the path integral quantization that leads to the island formula does not respect the commutativity of the left and right radiative algebras. 

While the existence of islands has not been investigated  for the CGHS model, an island computation for an evaporating black hole in the closely related RST model \cite{Russo:1992ax,Fiola:1994ir} was first carried out in \cite{Hartman:2020swn} and recovered the rising-and-falling Page curve. Island computations in dilatonic black hole settings have since been carried out in \cite{Gautason:2020tmk,Wang:2021mqq,Yu:2022xlh,Anegawa:2020ezn,Tian:2022pso,Fitkevich:2025zcf}.

We argue that the contradiction between commutativity and islands that we establish for the CGHS model should hold for the RST model. This points to a possible contradiction with \cite{Hartman:2020swn}. 

The version of HoI that we construct, on the other hand, does respect this commutativity. Thus our results suggest that the differences between the two approaches, in the context of CGHS (and possibly RST) model, go beyond the choice of algebra and stem from inequivalent quantizations of gravity.

The paper is organized as follows. In section~\ref{sec:CGHS}, we review the CGHS model
in some detail. Section~\ref{sec:phase_space} performs the classical analysis of
the CGHS model. The material in the first subsection reviews the covariant phase space analysis of the CGHS model. We then
obtain the radiative phase space, the Hamiltonian and the right cut charge. 
Section~\ref{sec:HoI} presents our main results, establishing HoI for the CGHS model
coupled to a massless scalar. In section~\ref{sec:algebras}, we discuss the tension between the two algebras. In our concluding section~\ref{sec:conclusion}, we summarize our results and discuss their implications. Useful details are collected in the appendices. 

 \section{CGHS Model and Evaporating Black Holes} \label{sec:CGHS}
This section reviews the CGHS model. 

 The action for the CGHS model coupled to a massless scalar is given by: 
\begin{equation} \label{action}
 S =\frac{1}{2} \int d^2x\, \sqrt{-g}\left( e^{-2\phi} \left(R + 4(\nabla\phi)^2 + 4\lambda^2\right) -\frac{1}{2} (\nabla f)^2 \right)   
\end{equation} 

where $\phi$ and $f$ are the dilaton and the matter scalar field respectively. 

As the CGHS model is 2-dimensional, the physical metric $g_{\mu\nu}$ is conformally flat. We thus fix a fiducial flat metric $\eta^{\mu \nu}$ and write \begin{equation}
    g^{\mu \nu}= \Omega(x) \eta^{\mu \nu}
\end{equation}
with $\Omega$ being the conformal factor. This is called conformal gauge.

For later convenience we set $\Phi=e^{-2\phi}$ and introduce a new field $\Xi=\Omega^{-1}\Phi$. Then our inverse metric reduces to $g^{\mu\nu}=\Xi^{-1}\Phi\eta^{\mu\nu}$. Then the entire geometry is completely determined by $\Xi$ and $\Phi$ and the field equations obtained by varying the action.

The equations of motion in conformal gauge are given by: 

\begin{align}
\Box_{(g)} f = 0 &\;\Longleftrightarrow\; \Box_{(\eta)} f = 0 \label{eom:matter}\\ \label{eom:dil}
\partial_+ \partial_- \Phi + \lambda^2 \Xi &= 0 \\ \label{eom:grav}
\Phi \, \partial_+ \partial_- \ln \Xi &= 0 
\end{align}

and the constraints

\begin{align}\label{con1}
-\partial_+^2 \Phi + \partial_+ \Phi \, \partial_+ \ln \Xi &= T_{++}\\
-\partial_-^2 \Phi + \partial_- \Phi \, \partial_- \ln \Xi &= T_{--}\label{con2}
\end{align}
Here $T_{\pm\pm}=\frac12(\partial_\pm f)^2$ are the non-vanishing
components of the stress-energy tensor of the matter field $f$ with
respect to $x^\pm=x^0\pm x^1$. In particular, $T_{+-}=0$.

From \eqref{eom:matter}, we see that the scalar $f$ is subject to the wave equation in fiducial flat space. The general solution is given by:
\begin{align*}
    f(x^{\pm})=f^{+}(x^{+}) +f^{-}(x^{-}) 
\end{align*}
where $f^{\pm}$ are arbitrary well-behaved functions of their arguments.

We restrict throughout to the collapse sector built on the linear dilaton
vacuum. The primordial mass and the remaining homogeneous dilaton data are
fixed to zero. Within this sector, the solutions for $\Xi$ and $\Phi$ for
any given regular $f$ were given in \cite{Kuchar:1996zm}. They are best
expressed by first introducing new coordinates:
$\lambda y^\pm=\pm e^{\pm \lambda x^\pm}$:

\begin{align}
\Xi(x^\pm) &= - \lambda^2 y^{+} y^{-} \\
\Phi(x^\pm) &= \Xi
- \frac12
\int_{0}^{y^{+}} d\bar{y}^{+}
\int_{0}^{\bar{y}^{+}} d\bar{\bar{y}}^{+}
\left(
\frac{d f^{+}}{d\bar{\bar{y}}^{+}}
\right)^2
-\frac12
\int_{0}^{y^{-}} d\bar{y}^{-}
\int_{0}^{\bar{y}^{-}} d\bar{\bar{y}}^{-}
\left(
\frac{d f^{-}}{d\bar{\bar{y}}^{-}}
\right)^2
\label{eq:Phi-solution}
\end{align}

We thus see that, in this sector, $f$ is the dynamical degree of freedom of
the CGHS model coupled to a matter scalar field, and the other fields are
determined in terms of it.

We consider a setup where the black hole is formed by a matter collapse of the left-moving modes, whereas the Hawking radiation is carried by the right-moving modes. The problem is extracting the state of the Hawking radiation from the right boundary, assuming that the black hole has fully evaporated and the singularity has been resolved. Thus our setting is the same as that of \cite{Ashtekar:2008jd} even though our methods are different.

It is useful to discuss the distinction between classical, semiclassical and quantum CGHS black holes. These differ in causal structure, fate of the singularity and the completeness of the two boundaries.

At the classical level, the dilaton $\Phi = e^{-2\phi}$ vanishes along a spacelike curve in the interior, producing a curvature singularity. There is no Hawking radiation and hence no evaporation. $\mathcal{I}^+_R$, is complete in the physical metric $g$ but not in the fiducial flat metric $\eta$ \cite{Giddings:1994pj,Strominger:1994tn}. The left boundary $\mathcal{I}^+_L$ is not complete---it is truncated by the singularity.

At the semiclassical level, one includes the one-loop quantum corrections, which introduce Hawking radiation and backreaction. The black hole radiates and shrinks. In this case the singularity is transient and $\mathcal{I}^+_R$ is no longer complete as it is terminated by a naked (``thunderbolt'') singularity. In the RST variant, however, there is no naked singularity and $\mathcal{I}^+_R$ is complete. The island computation of \cite{Hartman:2020swn} considers the semiclassical RST setting.

For the main HoI results,  we assume that the black hole fully evaporates and the singularity disappears. In this paper, we will make the same assumption. This implies that $\mathcal{I}^+_R,\,\mathcal{I}^+_L$ are complete in the fiducial flat metric $\eta$. That is, the affine parameter $x^-$ does not terminate at any finite value. This is the ``quantum'' setting which we will adopt in our proofs in Section \ref{sec:HoI}\,A and \ref{sec:HoI}\,B. As we will see in Sec.~IV\,C, state recovery is possible even without this assumption, that is in the presence of a horizon. However, the state recovered in this case can be a mixed state.

\section{Boundary Phase Space of CGHS model} \label{sec:phase_space}
\subsection{Covariant Phase Space Analysis of CGHS}

We start with a covariant phase space analysis \cite{Crnkovic:1986ex,Zuckerman:1989cx,Lee:1990nz,Barnich:1991tc,Wald:1993nt,Iyer:1994ys,Iyer:1995kg,Wald:1999wa,Barnich:2001jy} of CGHS model. The phase space of CGHS and related models, from both canonical and covariant approaches, have been studied in \cite{Navarro-Salas:1994cyz,Navarro-Salas:1994vfm,Kim:1995jta,Ruzziconi:2020wrb,Pedraza:2021cvx,McNees:2023tus}. Here, we briefly sketch the analysis to be self-contained.

Varying the Lagrangian $L$ defined via \eqref{action}, one obtains: 
\begin{equation}
\delta L
=
E_{\mu\nu}\,\delta g^{\mu\nu}
+
E_{\phi}\,\delta \phi
+
E_{f}\,\delta f
+
\nabla_\mu \Theta^\mu
\end{equation}

where $E_f$, $E_\phi$, and $E_{\mu\nu}$ are the Euler--Lagrange
expressions for the matter field, dilaton, and metric. The vector
$\Theta^\mu$ is the presymplectic potential.
 
\begin{equation}
\begin{aligned}
\Theta^\mu 
&= \frac{1}{2} \Big[
e^{-2\phi} \nabla^\mu (g_{\alpha\beta} \delta g^{\alpha\beta})
- e^{-2\phi} \nabla_\nu \delta g^{\mu\nu} - \nabla^\mu e^{-2\phi} \, g_{\alpha\beta} \delta g^{\alpha\beta}
+ \nabla_\nu e^{-2\phi} \, \delta g^{\mu\nu} + \frac{2 \nabla^\mu e^{-2\phi} \,\delta e^{-2\phi}}{e^{-2\phi}}
- \nabla^\mu f \,\delta f
\Big].\label{presymplectic}
\end{aligned}
\end{equation}
The details of the computation are given in Appendix \ref{appendixA}

 We now express the presymplectic potential in conformal gauge. Using:
\begin{equation}
\begin{aligned}
\delta g^{\mu\nu}&= \frac{\delta\Phi}{\Phi}\,g^{\mu\nu}\\
\nabla_\nu \delta g^{\mu\nu} &= \nabla^\mu\!\left(\frac{\delta\Phi}{\Phi}\right)\\
g_{\mu\nu}\delta g^{\mu\nu}&= \Xi\Phi^{-1}\eta_{\mu\nu}\frac{\delta\Phi}{\Xi}\,\eta^{\mu\nu}=2\frac{\delta\Phi}{\Phi}
\end{aligned}
\end{equation}
The residual conformal gauge is fixed so $\delta \Xi=0$ holds identically.

Substituting these relations into the expression for $\Theta^\mu$, we obtain
 
\begin{equation}
\Theta^{\mu}
= \frac{1}{2}
\left(
\nabla^{\mu} \delta \Phi
- \nabla^{\mu} f \, \delta f
\right).
\end{equation}
In conformal gauge with fixed $\Xi$, its density satisfies
$\sqrt{-g}\,\nabla^\mu\delta\Phi
=\delta(\sqrt{-g}\,\nabla^\mu\Phi)$. We can therefore drop the first
term because it is exact on phase space.
So the improved presymplectic potential is:

\begin{equation}\label{presym}
\Theta^\mu=-\frac{1}{2}\nabla^{\mu} f \, \delta f
\end{equation}

We now construct the symplectic current, which is defined as the
antisymmetrized variation of the presymplectic potential,
\begin{equation}\label{ometa}
\omega^\mu(\delta_1,\delta_2)
:= \delta_1 \left(\sqrt{-g}\Theta^\mu(\delta_2)\right)
- \delta_2 \left(\sqrt{-g} \Theta^\mu(\delta_1)\right)
\end{equation}
Here, $\delta_1$ and $\delta_2$ denote two independent variations on phase space.

From \eqref{presym}, we have that
\begin{equation}
\sqrt{-g} \Theta^\mu =-\frac{1}{2}\sqrt{-g}g^{\mu \nu}\partial_\nu f \, \delta f=-\frac{1}{2}\sqrt{-\eta}\eta^{\mu \nu}\partial_\nu f \delta f
\end{equation}

Substituting in \eqref{ometa}, we get:
\begin{equation}
\omega^\mu_{\rm density} 
= \frac{1}{2}\sqrt{-\eta}\;\eta^{\mu\nu}
\bigl[\delta_1 f\;\partial_\nu(\delta_2 f) - \delta_2 f\;\partial_\nu(\delta_1 f)\bigr]\,.
\end{equation}
The associated symplectic form is obtained by integrating $\omega^\mu$ over a complete Cauchy hypersurface $\Sigma$: 
\begin{equation}
\Omega[\delta_1,\delta_2]
= \int_{\Sigma} d\Sigma_\mu \, \omega^\mu(\delta_1,\delta_2),
\label{eq:symplectic_form}
\end{equation}
where $d\Sigma_\mu$ is the directed surface element on $\Sigma$.

This is then given by: 
\begin{equation} \label{fullo}
\Omega=\frac{1}{2}\int d\Sigma_{\mu}\,\sqrt{-\eta}\,\eta^{\mu\nu}
\left[
\delta_{1} f\,\partial_{\nu}(\delta_{2} f)
-
\delta_{2} f\,\partial_{\nu}(\delta_{1} f)
\right]
\end{equation}

 The vanishing of the
\(\nabla_\mu \delta\Phi\)
contribution to \eqref{fullo} reflects the absence of propagating gravitational degrees of freedom in two-dimensional dilaton gravity.
The dilaton field \(\Phi\) lies along the null directions of the presymplectic form. Thus, in the sector with fixed homogeneous dilaton data, the reduced phase space is parametrized by the matter field \(f\) alone, with the symplectic form \eqref{fullo}.

Following \cite{Kuchar:1996zm}, we take the phase space to consist of  field configurations with Schwartz-class falloff on the open spatial manifold $\mathbb{R}$:
\begin{equation}\label{eq:schwartz}
f(x),\;\pi_f(x) \;\in\; \mathcal{S}(\mathbb{R})\,,
\end{equation}
where $\pi_f=\dot f/2$ is the canonical momentum. This is a sufficient
falloff condition for the Klein--Gordon symplectic structure and the
matter variational principle on a noncompact spatial slice. A nonzero
constant Cauchy configuration $f=c$ is excluded. The condition is not,
however, an assumption that the undifferentiated chiral fields are
Schwartz functions. Writing $q(x)=f(0,x)$ and $u=x^-$, the wave equation
gives
\[
 \partial_- f^-(u)=\pi_f(-u)-\frac12 q'(-u),
 \qquad
 f^-(+\infty)-f^-(-\infty)
 =\int_{-\infty}^{\infty}dx\,\pi_f(x)=p_0.
\]
Thus $\partial_-f^-$ is Schwartz, whereas $f^-$ may have different finite
endpoint limits. Equations (29)--(31) account for these limits explicitly.
Nontrivial radiation is also possible for $p_0=0$: a nonzero chiral
derivative can have vanishing integral without having vanishing energy
flux.

For the asymptotic dilaton coefficients used below, we further restrict
to configurations with finite Kruskal moments:
\begin{equation}
\int_{-\infty}^{\infty}dx^{+}\,
e^{-\lambda x^{+}}(\partial_{+}f^{+})^{2}<\infty,
\qquad
\int_{-\infty}^{\infty}dx^{-}\,
e^{\lambda x^{-}}(\partial_{-}f^{-})^{2}<\infty.
\label{eq:kruskal-moments}
\end{equation}
This condition is stronger than Schwartz falloff. It holds for compactly
supported data and for data with suitable exponential falloff. It ensures
that the asymptotic coefficients defined below exist.

As shown in \cite{Kuchar:1996zm}, the symplectic form given by \eqref{fullo} is non-degenerate on the Schwartz space $\mathcal{S}(\mathbb{R})$.

\subsection{Radiative Phase Space of CGHS model}
\label{subsec:radiative-phase-space}
In this section, we construct the radiative phase space of the two-dimensional CGHS dilaton gravity model at $\mathcal{I}^+_R$ , obtain the Hamiltonian and study the asymptotic symmetries.

Radiative phase space was pioneered in \cite{Ashtekar:1981hw,Ashtekar:1981bq,Ashtekar:1981sf,Ashtekar:1987tt}. A particularly explicit modern reference is \cite{Campiglia:2015kxa}.

For the main complete-boundary analysis of Secs.~IV\,A and IV\,B,
we assume that the black hole forms and evaporates away in finite time. It is thus reasonable to restrict to solutions where the outgoing flux at $\mathcal{I}^+_R$ vanishes at very early and very late times. We therefore impose that 
\begin{equation}\label{fluxcond}
 \partial_-f^-|_{\mathcal{I}^+_{R,+}}
 =\partial_-f^-|_{\mathcal{I}^+_{R,-}}=0.
\end{equation}
This condition will be useful when we construct the Hamiltonian.

On a family of spacelike Cauchy slices $\Sigma_t$ defined by 
$t = \frac{1}{2}(x^+ + x^-)$, the symplectic form \eqref{fullo} is
\begin{equation}\label{eq:full-omega}
\Omega_{\Sigma_t}(\delta_1,\delta_2) 
= \frac{1}{2}\int_{\Sigma_t} dx\;
\bigl[\delta_1 f\;\partial_t(\delta_2 f) 
- \delta_2 f\;\partial_t(\delta_1 f)\bigr]\,,
\end{equation}
where $f = f^+(x^+) + f^-(x^-)$ and both chiralities are present. In the
future null limit, a complete Cauchy slice approaches both components of
future null infinity. In the absence of flux through any other boundary,
its symplectic form decomposes as $\Omega_{\Sigma}=\Omega_L+\Omega_R$.
We now select the right component. Equivalently, we take $t\to\infty$
with $x^-$ held fixed and pull back the symplectic current to
$\mathcal{I}^+_R$. The directed surface element then aligns with $dx^-$
and the contraction projects onto $\partial_-$, giving
\begin{equation}\label{eq:null-limit}
\Omega_R = \frac{1}{2}\int_{\mathcal{I}^+_R} dx^-\;
\bigl[\delta_1 f\;\partial_-(\delta_2 f) 
- \delta_2 f\;\partial_-(\delta_1 f)\bigr]\,.
\end{equation}
Since $f^+$ depends only on $x^+$, we have $\partial_- f^+ = 0$ identically. Therefore 
$\partial_-(\delta f) = \partial_-(\delta f^-)$, and \eqref{eq:null-limit} becomes
\begin{equation}\label{eq:split}
\Omega_R = \Omega_{f^-} + \Omega_{\rm cross}\,,
\end{equation}
where
\begin{equation}\label{eq:omega-minus}
\Omega_{f^-} = \frac{1}{2}\int_{\mathcal{I}^+_R} dx^-\;
\bigl[\delta_1 f^-\;\partial_-(\delta_2 f^-) 
- \delta_2 f^-\;\partial_-(\delta_1 f^-)\bigr]
\end{equation}
is the symplectic form of the right-moving sector, and
\begin{equation}\label{eq:cross}
\Omega_{\rm cross} = \frac{1}{2}\int_{\mathcal{I}^+_R} dx^-\;
\bigl[\delta_1 f^+\;\partial_-(\delta_2 f^-) 
- \delta_2 f^+\;\partial_-(\delta_1 f^-)\bigr]
\end{equation}
is a cross term coupling the two chiralities. Since $\delta f^+(x^+\to\infty)$ is constant 
along $\mathcal{I}^+_R$, it factors out of the integral:
\begin{equation}\label{eq:cross-boundary}
\Omega_{\rm cross} 
= \frac{1}{2}\,\delta_1 f^+(\infty)
\Bigl[\delta_2 f^-\Bigr]_{x^-=-\infty}^{x^-=+\infty}
- (1\leftrightarrow 2)\,.
\end{equation}
This is a boundary term involving only the zero mode of $f^-$---the net change in the field 
value across $\mathcal{I}^+_R$.

 We now show that the cross term (\ref{eq:cross-boundary}) vanishes on the phase space 
defined by the Schwartz falloff condition (\ref{eq:schwartz}).

The chiral field $f^-(x^-)$ is a fixed function of $x^-$ alone, determined by the initial 
data $(f,\pi_f)\in\mathcal{S}\times\mathcal{S}$ via the wave equation. Since
$f,\pi_f\in\mathcal{S}(\mathbb{R})$, it follows that $\partial_- f^-\in\mathcal{S}(\mathbb{R})$
as a function of $x^-$. In particular, $\partial_- f^-$ is integrable, so $f^-(x^-)$ 
approaches finite limits as $x^-\to\pm\infty$, and these limits are related by
\begin{equation}
f^-(+\infty) - f^-(-\infty) = \int_{-\infty}^{\infty} dx^-\;\partial_- f^- 
= p_0\,,
\end{equation}
where $p_0 = \int_{-\infty}^{\infty}\pi_f\,dx$ is the total momentum. We fix (by convention) $f^-(+\infty) = 0$, giving $f^-(-\infty) = -p_0$. Likewise,
$f^+(x^+)$ approaches the finite limit $f^+(\infty) = p_0$.

Both boundary values appearing in (\ref{eq:cross-boundary}) are therefore controlled by the 
single phase-space variable $p_0$:
\begin{equation}
\delta f^+(\infty) = \delta p_0\,,\qquad
\bigl[\delta f^-\bigr]_{-\infty}^{+\infty} 
= 0 - \bigl(-\delta p_0\bigr) = \delta p_0\,.
\end{equation}
Substituting into (\ref{eq:cross-boundary}):
\begin{equation}
\Omega_{\rm cross} 
= \frac{1}{2}\,\delta_1 p_0\,\delta_2 p_0
\;-\; (1\leftrightarrow 2)
= \frac{1}{2}\,\delta_1 p_0\,\delta_2 p_0
- \frac{1}{2}\,\delta_2 p_0\,\delta_1 p_0 = 0\,.
\end{equation}
The cross term vanishes because both factors are proportional to the same variable: 
$\delta p_0\wedge\delta p_0 = 0$.
  
Thus, on the Schwartz phase space, the right-boundary symplectic form
reduces exactly to the right-moving contribution:
\begin{equation}\label{eq:rad-result}
\Omega_R = \Omega_{f^-} 
= \frac{1}{2}\int_{\mathcal{I}^+_R} dx^-\;
\bigl[\delta_1 f^-\;\partial_-(\delta_2 f^-) 
- \delta_2 f^-\;\partial_-(\delta_1 f^-)\bigr]\,.
\end{equation}
The left-moving sector has dropped out: kinematically through $\partial_- f^+ = 0$ in the 
bulk, and through the Schwartz falloff condition at the boundary.

\subsection{Boundary Hamiltonian}
\label{subsec:boundary-hamiltonian}
 
We now derive the Hamiltonian directly from the boundary symplectic form at future null infinity.  
Starting from the boundary presymplectic potential
\begin{equation}
\Theta(\delta)
=
-\frac12 \int_{\mathcal{I}^+_R} dx^- \, \delta f \, \partial_- f ,
\end{equation}
the corresponding symplectic form is
\begin{equation}
\Omega_R(\delta_1,\delta_2)
=
\delta_1 \Theta(\delta_2)-\delta_2 \Theta(\delta_1)
=
\frac12 \int_{\mathcal{I}^+_R} dx^- \,
\Big(
\delta_1 f \, \partial_- \delta_2 f
-
\delta_2 f \, \partial_- \delta_1 f
\Big).
\label{sympbdry}
\end{equation}

Now consider diffeomorphisms that preserve $\mathcal{I}^+_R$, that is  reparametrizations of $x^-$:
\begin{equation}
x^- \to x^- + \epsilon(x^-)
\end{equation}
The corresponding infinitesimal diffeomorphism:
\begin{equation}\label{as}
\xi = \epsilon(x^-)\,\partial_- ,
\end{equation}
under which the scalar field transforms as
\begin{equation}\label{induc}
\delta_\epsilon f = \mathcal{L}_\xi f = \epsilon(x^-)\,\partial_- f .
\end{equation}
The corresponding generator $Q_\epsilon$ is defined by
\begin{equation}
\delta Q_\epsilon = \Omega_R(\delta_\epsilon,\delta).
\label{hamdef}
\end{equation}

Substituting \eqref{induc} into \eqref{sympbdry}, we obtain
\begin{align}
\delta Q_\epsilon
&=
\frac12 \int_{{\mathcal{I}}^+_{R}} dx^- \,
\Big[
\epsilon\,\partial_- f \, \partial_- (\delta f)
-
\delta f \, \partial_- \big(\epsilon\,\partial_- f\big)
\Big]
\nonumber\\
&=
\int_{{\mathcal{I}}^+_{R}} dx^- \,
\epsilon\,\partial_- f\,\partial_-(\delta f)
-
\frac12
\Big[
\epsilon\,\partial_- f\,\delta f
\Big]_{\mathcal{I}^+_{R,-}}^{\mathcal{I}^+_{R,+}} .
\end{align}

The bulk term is integrable:
\begin{equation}
\int_{{\mathcal{I}}^+_R} dx^- \, \epsilon\,\partial_- f\,\partial_-(\delta f)
=
\delta \left[
\frac12 \int_{{\mathcal{I}}^+_R} dx^- \, \epsilon(x^-)\,(\partial_- f)^2
\right].
\end{equation}
Therefore the variation of $Q_\epsilon$ is
\begin{equation}
\delta Q_\epsilon
=
\delta \left[
\frac12 \int_{\mathcal{I}^+_R} dx^- \, \epsilon(x^-)\,(\partial_- f)^2
\right]
-
\frac12
\Big[
\epsilon\,\partial_- f\,\delta f
\Big]_{\mathcal{I}^{+}_{R,-}}^{\mathcal{I}^{+}_{R,+}}.
\label{fullhamvar}
\end{equation}
The second term is a corner contribution from the two endpoints of $\mathcal{I}^+_R$.  It vanishes due to \eqref{fluxcond}. 
Then \eqref{fullhamvar} integrates to
\begin{equation}\label{fullham}
Q_\epsilon
=
\frac12 \int_{\mathcal{I}^+_{R}} dx^- \, \epsilon(x^-)\,(\partial_- f)^2
+
M_\epsilon =
 \int_{\mathcal{I}^+_{R}} dx^- \, \epsilon(x^-)\,T_{--}
+
M_\epsilon .
\end{equation}
where $M_{\epsilon}$ is constant on the right radiative phase space and is
not fixed by its symplectic form. For the radiative symmetry generators we
choose $M_{\epsilon}=0$. The representative selected by the asymptotic
dilaton is introduced separately below.

For $\epsilon=1$, the charge is the boundary Hamiltonian that generates rigid translations along $x^-$, which we will refer to as the right Hamiltonian:
\begin{equation}
\boxed{
H_R
=
 \int_{\mathcal{I}^+_R} dx^- \, T_{--}.
}
\label{Hfinal}
\end{equation}

\subsection{Right Cut Charge from the Asymptotic Dilaton Constraint}
\label{subsec:right-cut-charge}

We now construct the right cut charge which will be used in the next section to construct the asymptotic gravitational algebra\footnote{We thank Ronak Soni for pointing out an error in defining the right cut charge in the previous version}. We start by extracting the asymptotic dilaton coefficient from which the right cut charge
will be constructed. All coefficients below refer to the conformal gauge
and the asymptotic coordinates fixed above. Since $\Xi$ is fixed and
$g^{\mu\nu}=\Xi^{-1}\Phi\eta^{\mu\nu}$, the asymptotic expansion of
$\Phi$ encodes the geometry in this gauge. 

At large $y^{+}$ with $y^{-}$ fixed, the full dilaton has the expansion
\begin{equation}
\Phi\Big|_{\scri^{+}_{R}}
=-y^{+}A_{R}(y^{-})+F_{R}(y^{-})+o(1),
\qquad y^{+}\to\infty.
\label{eq:right-dilaton-expansion}
\end{equation}
The term denoted by $o(1)$ contains only contributions that vanish as
$y^+\to\infty$.
The coefficients are defined  by
\begin{equation}
A_{R}(y^{-})
=-\lim_{y^{+}\to\infty}\frac{\Phi}{y^{+}},
\qquad
F_{R}(y^{-})
=\lim_{y^{+}\to\infty}
\left(\Phi+y^{+}A_{R}(y^{-})\right).
\label{eq:right-coefficients}
\end{equation}

For the classical solution \eqref{eq:Phi-solution}, the leading
coefficient is
\begin{equation}
A_{R}(y^{-})=\lambda^{2}y^{-}+C_{+},
\label{eq:AR-onshell}
\end{equation}
where
\begin{equation}
C_{+}
=\frac12\int_{0}^{\infty}d\hat y^{+}\,
\left(\frac{df^{+}}{d\hat y^{+}}\right)^{2},
\label{eq:Cplus-matter}
\end{equation}
and
\begin{equation}
\begin{split}
F_{R}(y)
={}&\frac12\int_{0}^{\infty}d\hat y^{+}\,\hat y^{+}
\left(\frac{df^{+}}{d\hat y^{+}}\right)^{2}
\\
&-\frac12\int_{0}^{y^{-}}d\bar y^{-}
\int_{0}^{\bar y^{-}}d\hat y^{-}\,
\left(\frac{df^{-}}{d\hat y^{-}}\right)^{2}.
\end{split}
\label{eq:FR-matter}
\end{equation}
Equation \eqref{eq:FR-matter} is the on-shell matter representation of the the cut-local coefficient $F_R$.

For the HoI construction, we use $F_R$ to define the one-sided right-cut
charge :
\begin{equation}
\QB(u)=\lambda
\Big(F_{R}(y)-y\partial_{y}F_{R}(y)\Big)
\Big|_{y=y^{-}(u)},
\qquad
\lambda y^{-}(u)=-e^{-\lambda u}.
\label{eq:QB-def}
\end{equation}

Here locality at the cut means locality of the asymptotic germ. The
quantities $F_R$ and $\partial_uF_R$ are determined by the transverse
asymptotic expansion in an arbitrarily small neighbourhood of the cut. 

Since
\begin{equation}
\partial_{u}F_{R}=-\lambda y\partial_{y}F_{R},
\end{equation}
the charge can also be written as
\begin{equation}
\QB(u)=\partial_{u}F_{R}+\lambda F_{R}.
\label{eq:QB-u}
\end{equation}
At $\scri^{+}_{R}$ one has $\partial_u\ln\Xi=-\lambda$. Moreover,
$\partial_u y^{-}=-\lambda y^{-}$, so \eqref{eq:AR-onshell} gives
\begin{equation}
\left(\partial_u^2+\lambda\partial_u\right)A_R=0.
\label{eq:AR-constraint}
\end{equation}
The coefficient of $y^{+}$ in the $(--)$ constraint therefore vanishes.
Its finite part gives
\begin{equation}
-\partial_{u}^{2}F_{R}-\lambda\partial_{u}F_{R}=T_{--}(u).
\label{eq:constraint-at-scri}
\end{equation}
Combining \eqref{eq:QB-u} and \eqref{eq:constraint-at-scri}, we obtain
the balance law
\begin{equation}
\frac{d\QB}{du}=-T_{--}(u).
\label{eq:massloss}
\end{equation}

At the future corner of $\scri^{+}_{R}$, $y^{-}\to0^{-}$. Equations
\eqref{eq:QB-def} and \eqref{eq:FR-matter} give
\begin{align}
\QB(+\infty)
&=\lambda F_R(0)
\nonumber\\
&=\frac{\lambda}{2}\int_{0}^{\infty}d\hat y^{+}\,\hat y^{+}
\left(\frac{df^{+}}{d\hat y^{+}}\right)^{2}
\nonumber\\
&=\frac12\int_{-\infty}^{\infty}dx^{+}\,
(\partial_{+}f^{+})^{2}
=H_L.
\label{eq:QB-future-corner}
\end{align}
We define
\begin{equation}
\Htot\equiv\QB(-\infty)=H_{L}+H_{R},\qquad
H_{L}=\int_{\scri^{+}_{L}}dx^{+}\,T_{++},\qquad
H_{R}=\int_{\scri^{+}_{R}}dx^{-}\,T_{--}.
\label{eq:Htot}
\end{equation}
Integrating \eqref{eq:massloss} with the future-corner condition yields
\begin{equation}
\boxed{
\QB(u)=H_{L}+\int_{u}^{\infty}dx^{-}\,T_{--}(x^{-})
=\Htot-\int_{-\infty}^{u}dx^{-}\,T_{--}(x^{-}).
}
\label{eq:QB-flux}
\end{equation}
The constraint determines differences of the charge between cuts. By
itself it permits a cut-independent addition. The future-corner condition
fixes this ambiguity in the sector considered here. Thus $\QB$ is the
normalized Bondi charge used in our gravitational completion. We do not
claim that it is the only asymptotic gravitational observable at the cut.
As a normalization check, temporarily restoring a constant homogeneous
mass gives $\Phi=\Xi+M/\lambda$, and hence $F_R=M/\lambda$ and $\QB=M$.

The pure right-moving future flux is
\begin{equation}
M_{R}(u)
\equiv\int_{u}^{\infty}dx^{-}\,T_{--}(x^{-})
=\QB(u)-H_{L}.
\label{eq:MR-def}
\end{equation}
It is not determined by a single right cut, since the subtraction in
\eqref{eq:MR-def} requires the separate operator $H_L$. The invariant
Bondi mass associated with a complete cut is $M_{\mathrm{tot}}(u,v)$,
defined in \eqref{eq:Sigma-def}.

At the quantum level, the stress tensor and the asymptotic coefficients
are understood as renormalized operators. Since the stress tensor is an
operator-valued distribution, the half-line integrals are understood
with a smooth regulator at the cut.\footnote{For example,
$\int_u^\infty dx^-\,T_{--}(x^-)$ denotes
$\int_{-\infty}^{\infty}dx^-\,s_\delta(x^--u)T_{--}(x^-)$, where
$s_\delta(z)=0$ for $z\leq 0$, $s_\delta(z)=1$ for
$z\geq\delta$, and $s_\delta$ varies smoothly in between. The same
profile is translated to every cut. Its detailed form affects only a
small neighbourhood of the cut.}
We suppress this regulator in the notation. The same prescription is
used for the left and two-cut charges below.

We assume that the renormalized constraint and
\eqref{eq:QB-flux} continue to hold in the quantum theory. We also assume
that $H_L$ commutes with the right-moving sector. Normal ordering depends
on the choice of null coordinates. Changing between the $y^\pm$ and
affine coordinates adds a state-independent Schwarzian term, which is
included in the normalization of the charge. Such a term can shift the
zero of the charge, but it cannot remove the operator $H_L$. With affine
normal ordering, the corner charges
$\QB(-\infty)=\Htot$ and $\QB(+\infty)=H_L$ annihilate the global vacuum.
A finite-cut $\QB(u)$ generally does not.

\subsection{Asymptotic Symmetries}\label{sec:asymsym}

We have already identified diffeomorphisms that preserve $\mathcal{I}^+_R$ in \eqref{as}. It remains to verify that these transformations are genuine symmetries of the radiative phase space, i.e.\ that they preserve both the falloff conditions and the symplectic structure. In what follows we take $\epsilon(x)$ to belong to the class of multipliers of the Schwartz space that also preserve the domain defined by \eqref{eq:kruskal-moments}.

To check the first, note that they induce the transformation (using \eqref{induc}):
\begin{equation}\label{trn}
\delta_\epsilon(\partial_- f^-) = \epsilon'\,\partial_- f^- 
+ \epsilon\,\partial_-^2 f^-
\end{equation}

The RHS is a sum of products of smooth  functions of at most polynomial growth with Schwartz-class functions, 
and hence remains in $\mathcal{S}(\mathbb{R})$. 

For the latter, recall that a 
phase space vector field $\delta_\epsilon$ preserves the symplectic form if and only if it is Hamiltonian, i.e.\ $\iota_{\delta_\epsilon}\Omega_R = \delta Q_\epsilon$ for some 
phase space function $Q_\epsilon$. As we saw from \eqref{fullhamvar}, the variation $\delta Q_\epsilon$ is integrable and yields a well-defined charge 
$Q_\epsilon$ in \eqref{fullham}. The integrability of the charge guarantees $\mathcal{L}_{\delta_\epsilon}\Omega_R = d(\iota_{\delta_\epsilon}\Omega_R) = d(\delta Q_\epsilon) = 0$. This shows that the reparametrizations 
are canonical transformations on the radiative phase space.

We now determine the algebra satisfied by the generators $Q_\epsilon$. 
The Poisson brackets of the charges can be obtained by evaluating the variation of one charge under the transformation generated by another,
\begin{equation}
\delta_{\epsilon_2} Q_{\epsilon_1}
=
\{Q_{\epsilon_2}, Q_{\epsilon_1}\}.
\end{equation}

Using the expression for the charges derived in the previous section, the variation reduces to
\begin{equation}
\delta_{\epsilon_2} Q_{\epsilon_1}
= \int dx^- \, \epsilon_1 \, \delta_{\epsilon_2} T_{--}.
\end{equation}

The transformation of $T_{--}$ follows from $\delta_\epsilon f = \epsilon \partial_- f$ and is given by
\begin{equation}
\delta_\epsilon T_{--}
=
2 \epsilon' T_{--}
+
\epsilon \partial_- T_{--}.
\end{equation}

Substituting this into the variation of the charge, we obtain
\begin{equation}
\delta_{\epsilon_2} Q_{\epsilon_1}
=\int dx^- \, \epsilon_1
\left(
2 \epsilon_2' T_{--}
+
\epsilon_2 \partial_- T_{--}
\right)=
\int dx^-
(\epsilon_1 \epsilon_2' - \epsilon_2 \epsilon_1') \, T_{--}.
\end{equation}

We thus arrive at
\begin{equation}
\delta_{\epsilon_2} Q_{\epsilon_1}
=
Q_{\epsilon_1 \epsilon_2' - \epsilon_2 \epsilon_1'}.
\end{equation}
Therefore, the Poisson brackets of the generators satisfy
\begin{equation}
\{ Q_{\epsilon_2}, Q_{\epsilon_1} \}
=
Q_{\epsilon_1 \epsilon_2' - \epsilon_2 \epsilon_1'},
\end{equation}
With the convention
$\delta_{\epsilon_2}Q_{\epsilon_1}=\{Q_{\epsilon_2},Q_{\epsilon_1}\}$,
this reproduces the Lie bracket of vector fields
$\epsilon(x^-)\partial_-$. The corresponding active transformations at
fixed parameters form the associated anti-representation:
\begin{equation}
[\delta_{\epsilon_1}, \delta_{\epsilon_2}] f
=-
\delta_{(\epsilon_1 \epsilon_2' - \epsilon_2 \epsilon_1')}f.
\end{equation}

This is the classical Poisson algebra of reparametrizations of the null
coordinate $x^-$. In the quantum theory, normal ordering gives the
standard Virasoro central term. It vanishes on the translation and
dilation subalgebra.

A natural question is whether the asymptotic symmetry algebra of the CGHS model leads to a degenerate vacuum structure analogous to the supertranslation vacua in four dimensions. It does not, because as \eqref{induc} shows, the asymptotic symmetries act homogeneously on the fundamental field: the RHS of \eqref{induc} vanishes when $f^-=0$.
 
This leads to the corresponding charge $Q_\epsilon$ (defined via \eqref{fullham}) to be quadratic in $\partial_-f^-$. For a quadratic charge, the mode expansion takes the schematic form
\begin{equation}
  Q_\epsilon \;\sim\; \int dk\,dk'\;
  \widetilde{\epsilon}(k-k')\, a^\dagger_k\, a_{k'} 
  \;+\; \text{pair creation/annihilation terms}.
\end{equation}
That is, $Q_\epsilon$ contains no term linear in $a_k$ or
$a_k^\dagger$, and no $c$-number piece. For a generic non-affine
$\epsilon$, the pair-creation term can produce a two-particle excitation
when $Q_\epsilon$ acts on the Fock vacuum. The translation and dilation
generators annihilate the vacuum. In neither case does the action produce
a different vacuum with soft charge.
 This is in sharp contrast with four-dimensional asymptotically flat 
gravity, where the inhomogeneous action of BMS supertranslations on the shear tensor leads to an infinite family of degenerate soft vacua \cite{Ashtekar:1981hw,Ashtekar:1981bq,Ashtekar:1981sf,Strominger:2013jfa}. 

The matter action also has a global shift symmetry
\begin{equation}\label{eq:shift}
  f \;\longrightarrow\; f + c.
\end{equation}
But since constants are not in $\mathcal{S}(\mathbb{R})$, this transformation does not act on phase space. The associated charge is $p_0$, which is well-defined as a phase space function but does not generate any internal symmetry.
\section{Holography of Information and Information Loss in the CGHS Model}
\label{sec:HoI}

We will now establish holography of information for the for the gravitational completions of the CGHS radiative algebras defined below, in the sense of \cite{Laddha:2020kvp}. In each boundary setting, Result~1 is an
early-boundary algebra-completeness statement, and Result~3 follows
from it by isotony. Result~2 concerns the inclusion of later-cut
algebras in earlier-cut algebras; Result~4 is its finite-segment entropy
consequence. The finite-cut statements require the additional
assumptions identified below. The mathematical content of recovery is
equality of the specified early and global algebras, not merely equality
of their entropies. Thus agreement of two normal states on the early
algebra implies agreement on the corresponding global algebra. We
first treat both future null boundaries and then restrict to the right
boundary.

We assume that the physical Hilbert space factorizes as
\begin{equation}
\mathcal H=\Hfp\otimes\Hfm .
\label{eq:Hilbert-factorization-IV}
\end{equation}
We use this assumption to identify the global von Neumann algebras. An
algebraic version of the one-sided argument is given in
Appendix~\ref{app:algebraic}.

For intervals $I\subset\scri^{+}_{R}$ and $J\subset\scri^{+}_{L}$, define
the radiative algebras
\begin{align}
\Arad(I)
&=\mathrm{Alg}\{f^{-}(u),\partial_{-}f^{-}(u):u\in I\},
\nonumber\\
\Arad^{L}(J)
&=\mathrm{Alg}\{f^{+}(v),\partial_{+}f^{+}(v):v\in J\}.
\end{align}
The chiral scalar algebras satisfy isotony and additivity. Their global
representations are irreducible:
\begin{equation}
\Arad(\scri^{+}_{R})=B(\Hfm),
\qquad
\Arad^{L}(\scri^{+}_{L})=B(\Hfp).
\label{eq:global-radiative-algebras}
\end{equation}

Unlike four-dimensional asymptotically flat gravity, the CGHS model does not have an independent gravitational news. Hence the physical fall-off conditions do not automatically select a unique algebra of gravitational
observables at a cut. For the one-boundary case, we construct the gravitational algebra by appending  the right-cut charge $\QB$, which we extracted from the asymptotic dilaton, to $A_{\rm rad}$.

The reason for this choice is that the charge $\QB(u)$, like its 4-dimensional counterpart, 
obeys the Bondi balance law and generates translations of right radiative
observables to the future of the cut. So for the single boundary case, we define
\begin{equation}
\Agrav(I)
=
\mathrm{Alg}\{\Arad(I),\QB(u):u\in I\}.
\label{eq:Agrav-def}
\end{equation}
Note that while this is a geometrically motivated completion, it is not the unique
or maximal algebra of asymptotic gravitational observables. The charge
is local in the sense of the asymptotic germ explained in Sec.~III\,D;
its nonlocal on-shell expression in matter variables does not amount to
adjoining arbitrary global observables by definition. The physical content of our construction is that measurements of the radiation together with the specified charge extracted from the asymptotic dilaton suffice for the recovery statements below. This establishes HoI for a geometrically motivated gravitational completion of CGHS under the stated quantum assumptions. It does not require that this completion exhaust all observables available to a boundary observer. In the two-boundary
case we replace $\QB$ by $M_{\mathrm{tot}}$, defined below.

\subsection{Two-sided Holography of Information}
\label{subsec:two-sided-hoi}
The principal result of this subsection is Eq.~\eqref{eq:two-sided-result1}: the algebra near
the two past corners is $B(\mathcal H)$. The first proof uses the
past-corner normalization and Assumptions~1 and~2; it does not use the
exact translation Assumption~3. The alternative proof uses that
translation assumption. Result~3 follows from Result~1, whereas the
finite-cut Results~2 and~4 use Assumption~4.
A complete cut of the future boundary consists of two points, one at each boundary. We denote
them by $u\in\scri^{+}_{R}$ and $v\in\scri^{+}_{L}$. We assume that the
Kruskal moments are finite:
\begin{equation}
\int_{-\infty}^{\infty}dx^{+}\,
e^{-\lambda x^{+}}(\partial_{+}f^{+})^{2}<\infty,
\qquad
\int_{-\infty}^{\infty}dx^{-}\,
e^{\lambda x^{-}}(\partial_{-}f^{-})^{2}<\infty .
\label{eq:kruskal-falloff}
\end{equation}
This condition holds for compactly supported data and for data with
suitable exponential falloff. It ensures that the asymptotic coefficients
used below exist.

We already introduced the right cut charge $\QB(u)$ and its matter representation in
\eqref{eq:QB-def} and \eqref{eq:QB-flux}.
Now we introduce its left counterpart. Let $w=y^{+}(v)$. At
$\scri^{+}_{L}$, $x^-\to+\infty$ and hence $y^-\to0^-$. The $f^-$
double integral in \eqref{eq:Phi-solution} therefore vanishes at the
boundary without setting $f^-$ to zero. The boundary value of the full
dilaton is finite:
\begin{equation}
\Phi\big|_{\scri^{+}_{L}}
=F_{L}(w)=-G^{+}(w),
\qquad
G^{+}(w)=\frac12\int_{0}^{w}d\hat y\,(w-\hat y)N^{+}(\hat y),
\label{eq:left-dilaton-value}
\end{equation}
where
\begin{equation}
N^{+}(\hat y)=\left(\frac{df^{+}}{d\hat y}\right)^{2}.
\end{equation}
The corresponding cut charge is
\begin{equation}
\QBL(v)
=\lambda
\left(F_{L}-w\partial_{w}F_{L}\right)\Big|_{w=w(v)}
=\int_{-\infty}^{v}dx^{+}\,T_{++}(x^{+}).
\label{eq:left-charge}
\end{equation}
It obeys
\begin{equation}
\frac{d\QBL}{dv}=T_{++}(v),
\qquad
\QBL(-\infty)=0,
\qquad
\QBL(+\infty)=H_{L}.
\label{eq:left-charge-corners}
\end{equation}

The Bondi mass of the complete cut is the difference of the two right and left cut charges:
\begin{equation}
M_{\mathrm{tot}}(u,v)
\equiv\QB(u)-\QBL(v).
\label{eq:Sigma-def}
\end{equation}
On shell,
\begin{align}
M_{\mathrm{tot}}(u,v)
&=\int_{u}^{\infty}dx^{-}\,T_{--}(x^{-})
+\int_{v}^{\infty}dx^{+}\,T_{++}(x^{+})
\nonumber\\
&=\Htot
-\int_{-\infty}^{u}dx^{-}\,T_{--}(x^{-})
-\int_{-\infty}^{v}dx^{+}\,T_{++}(x^{+}).
\label{eq:Sigma-flux}
\end{align}
Thus $M_{\mathrm{tot}}(u,v)=M_R(u)+M_L(v)$, where $M_R$ and $M_L$
denote the future fluxes on the two boundary components.
It therefore satisfies
\begin{equation}
\frac{\partial M_{\mathrm{tot}}}{\partial u}=-T_{--}(u),
\qquad
\frac{\partial M_{\mathrm{tot}}}{\partial v}=-T_{++}(v).
\label{eq:Mtot-balance}
\end{equation}
Its corner values are
\begin{equation}
M_{\mathrm{tot}}(-\infty,-\infty)=\Htot,
\qquad
M_{\mathrm{tot}}(+\infty,+\infty)=0.
\label{eq:Mtot-corners}
\end{equation}

The two-cut mass is independent of the homogeneous ambiguity in the
dilaton solution. The linear terms are annihilated by the combinations
$F-\zeta F'$. A constant shifts $\QB$ and $\QBL$ equally. It therefore
cancels in \eqref{eq:Sigma-def}. Thus $M_{\mathrm{tot}}$ is the intrinsic
Bondi mass associated with a complete cut.

We define the gravitational algebra of a pair of intervals by
\begin{equation}
\Agravtwo(I;J)
=\mathrm{Alg}\left\{
\Arad(I),\Arad^{L}(J),
M_{\mathrm{tot}}(u,v):u\in I,\ v\in J
\right\}.
\label{eq:Agrav2}
\end{equation}
The finite-cut commutators are
\begin{align}
i[M_{\mathrm{tot}}(u,v),f^{-}(u')]
&=\Theta(u'-u)\,\partial_{-}f^{-}(u'),
\nonumber\\
i[M_{\mathrm{tot}}(u,v),f^{+}(v')]
&=\Theta(v'-v)\,\partial_{+}f^{+}(v').
\label{eq:Sigma-comm}
\end{align}
Thus the two-cut mass generates forward translations on both boundary
components.

For the first proof of Two-sided Result~1, we assume that the quantum
two-cut charge retains the past-corner normalization in
\eqref{eq:Mtot-corners}. Thus
$M_{\mathrm{tot}}(-\infty,-\infty)=\Htot$ belongs to the closure of the
algebra generated by charges in any neighbourhood of the two past
corners.

\begin{assumption}
\label{ass:positive-total}
The total Hamiltonian $\Htot$ has non-negative spectrum and a unique
ground state $|0\rangle=|0_{+}\rangle\otimes|0_{-}\rangle$, with
$\Htot|0\rangle=0$.
\end{assumption}

\begin{assumption}
\label{ass:radiative-analyticity}
The right and left radiative fields obey the usual positive-frequency
spectrum condition in their respective affine null coordinates. Their
vacuum matrix elements therefore have the standard positive-energy
analyticity property.
\end{assumption}

Assumption~\ref{ass:radiative-analyticity} is a condition on the
radiative representation. It does not require $\Htot$ to generate exact
translations of the radiative algebras in the full theory.

Let
\begin{equation}
I_{\epsilon}=J_{\epsilon}=(-\infty,-1/\epsilon),
\qquad \epsilon>0.
\label{eq:past-neighbourhoods}
\end{equation}

\textbf{Two-sided Result 1 (the full state is encoded near the two past
corners):} For every $\epsilon>0$,
\begin{equation}
\Agravtwo(I_{\epsilon};J_{\epsilon})
=B(\Hfp\otimes\Hfm)=B(\mathcal H).
\label{eq:two-sided-result1}
\end{equation}

\begin{proof}
Let
\begin{equation}
\mathcal R_{\epsilon}
=\Arad(I_{\epsilon})\vee\Arad^{L}(J_{\epsilon}).
\label{eq:two-sided-early-radiative}
\end{equation}
We first show that $\mathcal R_{\epsilon}|0\rangle$ is dense in
$\mathcal H$. Suppose that $|\Psi_{\perp}\rangle$ is orthogonal to all
such states. For any finite product of right and left radiative fields,
consider the matrix element
\begin{equation}
\kappa(\boldsymbol{u},\boldsymbol{v})
=\langle\Psi_{\perp}|X_{R}(\boldsymbol{u})
X_{L}(\boldsymbol{v})|0\rangle .
\label{eq:two-sided-correlator}
\end{equation}
It vanishes when every $u_i$ lies in $I_{\epsilon}$ and every $v_j$ lies
in $J_{\epsilon}$. This is a non-empty open set of insertion points.
Assumption~\ref{ass:radiative-analyticity} and the edge-of-the-wedge
argument therefore imply that \eqref{eq:two-sided-correlator} vanishes
identically.
The joined global radiative algebra acts irreducibly by
\eqref{eq:global-radiative-algebras}. Hence
$|\Psi_{\perp}\rangle=0$, proving the density statement.

The limit $\Htot=M_{\mathrm{tot}}(-\infty,-\infty)$ belongs to the von
Neumann closure of $\Agravtwo(I_{\epsilon};J_{\epsilon})$. Hence the
vacuum projector
\begin{equation}
P_{0}=|0\rangle\langle0|
=\lim_{\alpha\to\infty}e^{-\alpha\Htot}
\end{equation}
also belongs to this algebra. If $X_{n}|0\rangle$ is a dense family of
local states, the operators $X_{n}P_{0}X_{m}^{\dagger}$ generate the
finite-rank operators. Their von Neumann closure is $B(\mathcal H)$.
\end{proof}

We next introduce a stronger assumption. It gives a shorter proof of the
same result.

\begin{assumption}
\label{ass:total-translations}
The total Hamiltonian generates rigid translations of both radiative
algebras.
\end{assumption}

\begin{proof}[Alternative proof of Two-sided Result~1]
The past-corner relation \eqref{eq:Mtot-corners} implies that $\Htot$
belongs to the closure of
$\Agravtwo(I_{\epsilon};J_{\epsilon})$. The corresponding unitaries
therefore belong to this algebra. Assumption~\ref{ass:total-translations}
gives
\begin{align}
e^{is\Htot}\Arad(I_{\epsilon})e^{-is\Htot}
&=\Arad(I_{\epsilon}+s),
\nonumber\\
e^{is\Htot}\Arad^{L}(J_{\epsilon})e^{-is\Htot}
&=\Arad^{L}(J_{\epsilon}+s).
\label{eq:two-sided-rigid-translations}
\end{align}
Every finite interval on either boundary is contained in a suitable
translate of $I_{\epsilon}$ or $J_{\epsilon}$. The two translations may
be performed separately inside the same algebra. Isotony and additivity
then give
\begin{equation}
\Arad(\scri^{+}_{R})\vee\Arad^{L}(\scri^{+}_{L})
\subset\Agravtwo(I_{\epsilon};J_{\epsilon}).
\end{equation}
Equation \eqref{eq:global-radiative-algebras} identifies the left-hand
side with $B(\mathcal H)$. The reverse inclusion is immediate.
\end{proof}

Thus the complete initial state can be recovered from arbitrarily small
neighbourhoods of the two past corners. The first proof uses the
past-corner relation, positive-energy analyticity, and the unique ground
state. The alternative proof instead uses the stronger translation action
of $\Htot$.

The remaining two-sided results compare finite cuts. For these results we
make the stronger finite-cut assumption.

\begin{assumption}
\label{ass:two-cut-relations}
The commutators \eqref{eq:Sigma-comm} and the balance law
\eqref{eq:Mtot-balance} hold in the full theory, up to local operators at
the cuts. Each two-cut mass has a self-adjoint realization. For radiative
fields smeared away from the cuts, \eqref{eq:Sigma-comm} exponentiates to
the corresponding finite translations.
\end{assumption}

For fixed $\epsilon>0$, let
\begin{equation}
I_{u}=(u,u+\epsilon),
\qquad
J_{v}=(v,v+\epsilon).
\label{eq:small-cut-intervals}
\end{equation}

\textbf{Two-sided Result 2 (earlier pairs of cuts contain later
information):} If $u_{2}\leq u_{1}$ and $v_{2}\leq v_{1}$, then
\begin{equation}
\Agravtwo(I_{u_{1}};J_{v_{1}})
\subset
\Agravtwo(I_{u_{2}};J_{v_{2}}).
\label{eq:two-sided-monotonicity}
\end{equation}

\begin{proof}
By Assumption~\ref{ass:two-cut-relations}, the commutators
\eqref{eq:Sigma-comm} translate the right and left
radiative fields towards the future. Acting on a right-localized operator
translates only that operator, and similarly on the left. The two later
radiative algebras can therefore be generated independently from the
earlier intervals. The later masses follow from
\begin{align}
M_{\mathrm{tot}}(u_{1},v_{1})
&=M_{\mathrm{tot}}(u_{2},v_{2})
-\int_{u_{2}}^{u_{1}}du\,T_{--}(u)
\nonumber\\
&\quad
-\int_{v_{2}}^{v_{1}}dv\,T_{++}(v).
\label{eq:Mtot-integrated-balance}
\end{align}
The flux operators are obtained by translating local stress tensors from
the earlier intervals.
\end{proof}

The result also applies when one cut is held fixed.

Let
\begin{equation}
I_{<u}=(-\infty,u),
\qquad
J_{<v}=(-\infty,v).
\end{equation}

\textbf{Two-sided Result 3 (the past-segment entropy is independent of
both cuts):} For any normal state on $\mathcal H$, the entropy of its
restriction to $\Agravtwo(I_{<u};J_{<v})$ is independent of $u$ and $v$.

\begin{proof}
Choose $\epsilon$ such that $I_{\epsilon}\subset I_{<u}$ and
$J_{\epsilon}\subset J_{<v}$. By isotony and Two-sided Result 1,
\begin{equation}
\Agravtwo(I_{<u};J_{<v})=B(\mathcal H).
\end{equation}
The restricted state is therefore the same state on the same algebra for
every pair $(u,v)$.  Since the restricted state is the same state on the same algebra for every cut, it follows that its entropy is also cut-independent.
\end{proof}
 Note that the definition of entropy for a continuum boundary
segment requires a regulator and subtraction prescription. In the above statement and all subsequent statements about entropy, it is to be understood that we adopt a convention where the prescription is independent of the cut.

Finally, for a pure initial state, the two-sided Page curve is flat at zero.

\textbf{Two-sided Result 4 (finite-segment entropy):} With fixed lower cuts $(u_{1},v_{1})$, the entropy associated with
$\Agravtwo((u_{1},u_{2});(v_{1},v_{2}))$ is independent of the upper cuts
$u_{2}$ and $v_{2}$. Isotony gives one inclusion when an upper cut is
advanced. Two-sided Result 2 gives the reverse inclusion.
\subsection{One-sided Holography of Information}
\label{subsec:one-sided-hoi}
The target is now Eq.~\eqref{eq:result1}: the early right algebra equals the global
right completion, not $B(\mathcal H)$. Its left-sector content is only
$\mathcal C_L=\{H_L\}^{\prime\prime}$, as in Eq.~(98).
For Results~1 and~3, the proof below uses the past-corner normalization,
the stated operator charge relations, and Assumption~3 restricted to
the right radiative algebra. It does not use the finite-cut
Assumption~5. Results~2 and~4 use Assumption~5. Appendix B gives a
factorization-independent alternative using the stronger finite-cut
input. We recall the right completion:

\begin{equation}
\Agrav(I)
=\mathrm{Alg}\{\Arad(I),\QB(u):u\in I\}.
\end{equation}

Let
\begin{equation}
\mathcal C_{L}\equiv\{H_{L}\}''
\label{eq:left-energy-algebra}
\end{equation}
be the von Neumann algebra generated by the left Hamiltonian. It commutes
with the complete right radiative algebra. Equation
\eqref{eq:QB-flux} gives
\begin{equation}
\Agrav(\scri^{+}_{R})
=B(\Hfm)\,\bar\otimes\,\mathcal C_{L}.
\label{eq:global-grav}
\end{equation}
Indeed, every $\QB(u)$ is affiliated to the right-hand side. Conversely,
for any finite $q$, $M_R(q)$ is affiliated to the global right radiative
algebra and
\begin{equation}
H_{L}=\QB(q)-M_R(q).
\label{eq:HL-from-QB}
\end{equation}
The strong-commutativity assumption stated in
Section~\ref{app:algebraic} gives
\begin{equation}
e^{itH_L}=e^{it\QB(q)}e^{-itM_R(q)}.
\label{eq:HL-unitary}
\end{equation}

The projector proof of Two-sided Result~1 does not directly go through for the analogous
one-sided statement. Even under Assumptions~\ref{ass:positive-total} and
\ref{ass:radiative-analyticity}, the past-corner charge gives the global
vacuum projector
$P_0=P_{0,L}\otimes P_{0,R}$. Sandwiching it between right-local
operators produces operators of the form
$P_{0,L}\otimes X_RP_{0,R}Y_R^\dagger$. These generate only the
left-vacuum block. They do not generate
$\mathbf 1_{L}\otimes B(\Hfm)$ or recover $\mathcal C_L$. The argument
can be repeated after fixing a spectral sector of $H_L$, but assembling
the sectors already requires control of $\mathcal C_L$. We therefore do
not use the projector route.

For the first one-sided result we use the stronger
Assumption~\ref{ass:total-translations}, introduced in the two-sided
subsection, restricted to the right radiative algebra. We also use the
past-corner normalization in \eqref{eq:Htot} in the same operator sense
as in the two-sided construction. Thus $\QB(-\infty)=\Htot$ belongs to
the closure of the algebra generated by the charges in every
$I_\epsilon$.

\textbf{One-sided Result 1 (the right state and left energy are encoded
near the right past corner):} For every $\epsilon>0$,
\begin{equation}
\Agrav(I_{\epsilon})
=B(\Hfm)\,\bar\otimes\,\mathcal C_{L}
=\Agrav(\scri^{+}_{R}).
\label{eq:result1}
\end{equation}

\begin{proof}
The past-corner relation implies that the unitaries generated by $\Htot$
belong to the closure of $\Agrav(I_\epsilon)$. By
Assumption~\ref{ass:total-translations},
\begin{equation}
e^{is\Htot}\Arad(I_\epsilon)e^{-is\Htot}
=\Arad(I_\epsilon+s).
\label{eq:one-sided-rigid-translations}
\end{equation}
Every finite interval on $\scri^{+}_{R}$ is contained in a suitable
translate of $I_\epsilon$. Isotony and additivity therefore give
\begin{equation}
B(\Hfm)=\Arad(\scri^{+}_{R})
\subset\Agrav(I_{\epsilon}).
\end{equation}
Choose a finite $q\in I_{\epsilon}$. Then $\QB(q)$ is affiliated to
$\Agrav(I_{\epsilon})$. The operator $M_R(q)$ is affiliated to the same
algebra because the global right radiative algebra is already contained in
it. Equation \eqref{eq:HL-unitary} therefore gives
$\mathcal C_L\subset\Agrav(I_{\epsilon})$. This proves one
inclusion in \eqref{eq:result1}. The reverse inclusion follows from
\eqref{eq:QB-flux}.
\end{proof}

The one-sided algebra contains left data only through $\mathcal C_L$. It
can retain correlations between right observables and the left-energy
sectors, but it contains no left observables outside $\mathcal C_L$.

We now compare finite cuts. This requires a stronger, one-sided
finite-cut assumption. It is the counterpart of
Assumption~\ref{ass:two-cut-relations} and is not used in the proof of
One-sided Result~1. The cut charge has the commutator
\begin{equation}
i[\QB(u),f^{-}(u')]
=\Theta(u'-u)\,\partial_{-}f^{-}(u').
\label{eq:QB-comm}
\end{equation}
The left Hamiltonian makes no contribution to this commutator.

\begin{assumption}
\label{ass:right-cut-relations}
The renormalized commutator \eqref{eq:QB-comm} and the balance law
\eqref{eq:massloss} hold in the full theory, with any cut-local
counterterms included in the definition of the charge. Each cut charge
has a self-adjoint realization. On radiative fields smeared away from the
cut, \eqref{eq:QB-comm} exponentiates to the corresponding finite
translation.
\end{assumption}

We continue to use the intervals $I_u$ defined in
\eqref{eq:small-cut-intervals}.

\textbf{One-sided Result 2 (earlier right cuts contain later
information):} If $u_{2}<u_{1}$, then
\begin{equation}
\Agrav(I_{u_{1}})\subset\Agrav(I_{u_{2}}).
\label{eq:one-sided-monotonicity}
\end{equation}

\begin{proof}
Fix $q_1\in I_{u_1}$ and choose $q_2\in I_{u_2}$ with $q_2<q_1$. By
Assumption~\ref{ass:right-cut-relations}, the charge $\QB(q_2)$
translates radiative operators to the future of $q_2$. It therefore
generates the radiative algebra near $q_1$ from the algebra in
$I_{u_2}$. The later charge follows from
\begin{equation}
\QB(q_{1})
=\QB(q_{2})-\int_{q_{2}}^{q_{1}}du\,T_{--}(u).
\label{eq:one-sided-integrated-balance}
\end{equation}
The stress tensors in the integral are obtained by translating local
stress tensors from $I_{u_2}$. Since $q_1$ was arbitrary, every generator
of $\Agrav(I_{u_1})$ belongs to $\Agrav(I_{u_2})$.
\end{proof}

The next result uses One-sided Result~1 and therefore inherits
Assumption~\ref{ass:total-translations} from the two-sided subsection. It
does not require the finite-cut Assumption~\ref{ass:right-cut-relations}.

\textbf{One-sided Result 3 (the right past-segment state is independent of
the cut):} For any normal state, its restriction to $\Agrav(I_{<u})$ is
independent of $u$.

\begin{proof}
Choose $\epsilon$ such that $I_{\epsilon}\subset I_{<u}$. Isotony and
One-sided Result 1 give
\begin{equation}
\Agrav(I_{<u})=\Agrav(\scri^{+}_{R})
\end{equation}
for every $u$. The restricted state is therefore independent of the cut. As in the two-sided case, it follows that the entropy\footnote{In the one-sided case, this requires an additional point of convention. The abelian algebra $\mathcal C_L=\{H_L\}''$ registers only the probability
distribution of the left energy. With a discrete regulator, the entropy
of the one-sided algebra contains the Shannon entropy of this distribution
together with the average entropy of the right-moving state conditioned
on $H_L$. One may instead work in a fixed-$H_L$ sector, in which case the
Shannon term is absent. For a continuous spectrum, a common choice of
measure or energy binning is required. By convention, we keep this choice independent of
the cut.} is also independent of the cut.
\end{proof}

\textbf{One-sided Result 4 (finite-segment entropy):} With the same fixed
entropy convention and fixed $u_{1}$, the entropy associated with
$\Agrav((u_{1},u_{2}))$ is independent of $u_{2}$. Isotony gives one
inclusion when $u_2$ is advanced. One-sided Result 2 gives the reverse
inclusion.

\subsection{Implications of the HoI Results}

We note three implications.
\begin{itemize}

\item{Two-sided Page curve:}
If all horizons and singularities have evaporated with the black hole and both future null boundaries are complete, the full initial state can be
recovered from arbitrarily small neighbourhoods of the two past corners.
For a pure initial state, the two-sided Page curve would be flat at zero.

\item{One-sided Page curve:}
From the result about one-sided entropy being independent of the upper cut, it follows that the corresponding Page curve will be flat. This only requires that the right boundary is complete. The entropy value in this case
need not vanish for a pure state on $\mathcal H$. The right gravitational algebra contains only the probabilities of the left-energy sectors
and their correlations with right observables---it has no information of the
noncommuting left observables that distinguish states within an energy
sector or of the relative phases between sectors. This missing left
information can purify the restricted state. The restriction of a pure
global state to the right algebra can therefore be mixed and have
nonzero entropy.

\item{Information localization in the presence of a horizon:}
We now relax the fiducial completeness assumption used above. Suppose
the accessible physical right boundary has fiducial coordinate range
$(-\infty,x_H^-)$. A finite endpoint in this coordinate is not itself a
statement of incompleteness in the physical metric. The following
extension is conditional on the accessible radiative algebra, charge
decomposition, and restricted translation action stated here; these
are not consequences of the complete-evaporation setup. Suppose that
the physical right-cut charge admits the decomposition

\begin{equation}
\QB^{\mathrm{phys}}(u)
=H_{L}+\int_{u}^{x^{-}_{H}}dx^{-}\,T_{--}(x^{-})
+M_{\mathrm{trans}}.
\label{eq:physical-cut-charge}
\end{equation}
Here $M_{\mathrm{trans}}$ denotes the cut-independent contribution
carried by right-moving modes that do not reach the physical right
boundary. Let
$\mathcal H_{<x^{-}_{H}}$ be the Hilbert space of the radiative modes that
do reach this boundary. We assume that their global algebra is
$B(\mathcal H_{<x^{-}_{H}})$ and that $M_{\mathrm{trans}}$ commutes with
it.

The recovery argument is the same as in One-sided Result~1. At the past
corner,
\begin{equation}
\QB^{\mathrm{phys}}(-\infty)
=H_L+\int_{-\infty}^{x^{-}_{H}}dx^-\,T_{--}(x^-)
+M_{\mathrm{trans}}
=\Htot.
\end{equation}
We assume the physical-boundary analogue of
Assumption~\ref{ass:total-translations}: $\Htot$ generates rigid
translations of accessible radiative operators as long as their support
remains in $(-\infty,x^{-}_{H})$. An early radiative interval can then be
translated through every finite interval in this physical component, but
not beyond its endpoint. Isotony and additivity recover
$B(\mathcal H_{<x^{-}_{H}})$.

For any finite $q<x^{-}_{H}$, the accessible future flux
\begin{equation}
M_R^{\mathrm{phys}}(q)
=\int_q^{x^{-}_{H}}dx^-\,T_{--}(x^-)
\end{equation}
belongs to this radiative algebra. Equation
\eqref{eq:physical-cut-charge} then gives
\begin{equation}
\QB^{\mathrm{phys}}(q)-M_R^{\mathrm{phys}}(q)
=H_L+M_{\mathrm{trans}}.
\end{equation}
Let $\mathcal C_{\mathrm{ext}}
=\{H_{L}+M_{\mathrm{trans}}\}''$. For every $I_\epsilon$ contained in
the physical boundary, we obtain
\begin{equation}
\Agrav^{\mathrm{phys}}(I_{\epsilon})
=\Agrav^{\mathrm{phys}}((-\infty,x^{-}_{H}))
=B(\mathcal H_{<x^{-}_{H}})\,\bar\otimes\,
\mathcal C_{\mathrm{ext}}.
\label{eq:physAlgebra}
\end{equation}
Thus the accessible state can be recovered
from a neighbourhood of the past endpoint: the restriction of the state
to the algebra in \eqref{eq:physAlgebra} is completely determined there.
Only the combined charge $H_L+M_{\mathrm{trans}}$ is retained from the
inaccessible degrees of freedom. Noncommuting trans-horizon observables
and information within its charge sectors are absent. Coherences between
different sectors are also invisible to this abelian algebra. The full
trans-horizon state therefore cannot be reconstructed.

The entropy of the accessible algebra is independent of the cut because
both the algebra and the restricted state are independent of the cut. It is nonzero when
the accessible radiation is entangled with trans-horizon degrees of
freedom, as in a Hawking pair state.

The same argument applies in the two-sided case. Access to
$\scri^{+}_{L}$ recovers $H_L$ separately, leaving
$\{M_{\mathrm{trans}}\}''$ as the residual abelian datum while the
noncommuting trans-horizon algebra remains inaccessible.

Thus we see that information localization and information loss can be logically independent---even when information is lost, the accessible information can still be localized in an arbitrarily small neighbourhood. 

\end{itemize}

  \section{Does island entropy correspond to $A_{\rm rad}(\mathcal{I}_R)$?}\label{sec:algebras}

We now turn to the question of whether an island entropy can be identified
with the von Neumann entropy of the radiative algebra at the right boundary.
We first give the algebraic argument for CGHS and then discuss its
extension to the RST model.

\subsection{The Algebraic Argument}

From the phase-space analysis of the full coupled CGHS gravity–matter theory, we saw that the resulting right-boundary radiative phase space consists entirely of \(f^-\) modes and their derivatives. The radiative algebra \(\mathcal A_{\mathrm{rad}}(\mathcal I_R^+)\) therefore contains no independent \(f^+\) operators. Motivated by this commutativity in the classical theory, we consider quantum theories in which the left and right radiative algebras commute exactly.

Recovery of the full left-moving quantum state
is therefore impossible from the right gravitational completion studied
here.

The obstruction in the radiative algebra requires only exact commutativity
of $\mathcal A_{\rm rad}(\mathcal I_R^+)$ and
$\mathcal A_{\rm rad}(\mathcal I_L^+)$.
Suppose that island reconstruction applies to CGHS and that the island
contains left-moving observables represented in
$\mathcal A_{\rm rad}(\mathcal I_L^+)$. Choose two such Hermitian
operators $A_L,B_L$ whose commutator has a nonzero matrix element between
states $|\phi\rangle,|\psi\rangle$ in the code subspace
$\mathcal H_{\rm code}$ where reconstruction is claimed:
\begin{equation}
\langle\phi|[A_L,B_L]|\psi\rangle\ne 0.
\label{eq:island-left-commutator}
\end{equation}
A Hermitian reconstruction $A_R\in\mathcal A_{\rm rad}(\mathcal I_R^+)$
would reproduce the action of $A_L$ on this subspace,
\begin{equation}
A_R|\chi\rangle=A_L|\chi\rangle,
\qquad |\chi\rangle\in\mathcal H_{\rm code}.
\label{eq:island-reconstruction-action}
\end{equation}
Using this equality on the bra in the first term of the commutator and
on the ket in the second gives
\begin{align}
\langle\phi|[A_L,B_L]|\psi\rangle
&=\langle A_R\phi|B_L\psi\rangle
 -\langle B_L\phi|A_R\psi\rangle \nonumber\\
&=\langle\phi|[A_R,B_L]|\psi\rangle=0,
\label{eq:island-commutator-obstruction}
\end{align}
where the last equality follows from left/right commutativity.
This contradicts Eq.~\eqref{eq:island-left-commutator}. Neither
Hilbert-space factorization nor preservation of the code subspace by
$A_L$ and $B_L$ is required. Thus exact radiative commutativity is
incompatible with reconstructing this noncommuting left-moving algebra
in the right radiation.

There are then two possibilities for CGHS. Either islands with this
reconstruction interpretation do not exist, or they exist in
UV completions which violate the commutativity of the left and right
radiative algebras.  

While the relevant global island saddle has not been established for
CGHS, islands have been argued to exist for the closely related RST
model by Hartman et al.
\cite{Hartman:2020swn}. They applied the island prescription to an
evaporating RST black hole and obtained a rising-and-falling Page curve.
They also gave a local gravitational replica analysis near the QES,
using the eternal RST geometry, which reproduces the QES condition.
This supplies a path-integral argument for the prescription, but not a
construction of the global replica saddle for the evaporating black hole.
It therefore provides a concrete setting in which to examine the two
possibilities above.

The CGHS proof does not apply directly to RST: its gravitational dynamics
and strong-coupling boundary conditions differ, and we have not constructed
its asymptotic algebras here. Nevertheless, we argue that these differences
do not remove the obstruction for the infalling modes whose recovery is
predicted by the island prescription. The RST counterterm changes the
gravitational equations but leaves the minimally coupled free matter
action unchanged. The local matter field still separates into ingoing
and outgoing chiral modes. In the semiclassical island construction,
the island and the collected radiation lie in disjoint regions of a
common Cauchy slice \cite{Hartman:2020swn}. Local infalling matter
observables and the radiation observables therefore commute in this
semiclassical matter description. In particular, smeared chiral
derivatives inside the island supply a noncommuting infalling algebra
which commutes with the radiation algebra.

The reflecting boundary conditions require separate consideration,
since they can relate ingoing and outgoing modes. However, in the
post-Page scrambling experiment of \cite{Hartman:2020swn}, sufficiently
before the evaporation endpoint, the infalling excitation encounters
the spacelike segment of the strong-coupling boundary. It cannot reflect
there and return to the radiation. The island prescription nevertheless
predicts recovery on a scrambling-time scale, at the retarded time
associated with the excitation crossing the QES curve. Thus ordinary
reflection does not make the relevant infalling operators part of the
collected radiation algebra. The change in gravitational dynamics and
the reflecting boundary conditions do not, by themselves, evade the
obstruction. If a quantum completion retains the commutativity of these
infalling observables with the collected radiation, the same calculation
in Eq.~\eqref{eq:island-commutator-obstruction} rules out their
reconstruction there. For RST, this is the relevant commutativity
statement; it does not require identifying its global boundary algebras
with the two CGHS future-boundary algebras.

If this extension of the argument holds, there are likewise two
possibilities for RST. Either the proposed islands are not realized,
in which case a global replica calculation must fail to produce a
contributing saddle with the stated reconstruction interpretation, or
the path-integral quantization realizing them does not preserve exact
commutativity between the relevant infalling and radiation algebras.
A global replica derivation for the evaporating RST black hole remains
unavailable. The global solution of the replica deformation obtained in
\cite{Arias:2026zgz} is for the eternal RST black hole, to first order near
$n=1$, and does not settle this evaporating problem.

Presumably, in the second possibility, the path-integral quantization
provides some mechanism, such as topology change, that allows left-mover information
to end up in the right radiation. This is somewhat reminiscent of the
factorization problem
\cite{Penington:2019kki,Almheiri:2019qdq,Marolf:2020xie,Giddings:2020yes,Marolf:2020rpm,Engelhardt:2020qpv,Saad:2021uzi,Saad:2021rcu,Peng:2021vhs,Blommaert:2021fob,Blommaert:2022ucs,Benini:2022hzx,Schlenker:2022dyo,Colafranceschi:2023moh,Hernandez-Cuenca:2024pey,Boruch:2024kvv},
where wormhole correlations in the gravitational path integral challenge
the expected tensor product description of disconnected boundaries.

\subsection{Difference between Island and HoI Results}

In light of these results, the source of disagreement between the results
of island and HoI approaches needs to be re-examined. In
\cite{Antonini:2025sur,Geng:2026asi}, the difference was attributed to the
choice of algebra: $\mathcal{A}_{\mathrm{rad}}$ for islands versus
$\mathcal{A}_{\mathrm{grav}}$ for HoI\footnote{A complementary line of
work argues that the gravitational Gauss law of massless long-range
gravity obstructs island reconstruction
\cite{Geng:2020qvw,Geng:2021hlu,Geng:2023zhq,Geng:2025rov,Geng:2025bcb}.
In our case there is no coupling with a bath.}.

Our analysis suggests that the choice of quantization also plays a key
role. Under the assumptions of Sec.~\ref{sec:HoI}, HoI holds for our
gravitational completion in quantum theories where
$\mathcal A_{\rm rad}(\mathcal I_R^+)$ and
$\mathcal A_{\rm rad}(\mathcal I_L^+)$ commute exactly. This commutativity
is motivated by the reduced phase-space structure of the full coupled
CGHS theory derived in Sec.~\ref{sec:phase_space}. The argument above
shows that islands with the stated reconstruction interpretation cannot
exist in such a theory. Thus HoI can hold while islands are absent.

It was conjectured in \cite{Harlow:2020bee,Harlow:2021lpu} that the
Euclidean path integral in a gravitational effective field theory
correctly computes von Neumann entropies if and only if its
quantum-mechanical UV completion is holographic\footnote{The
contradiction between islands and global symmetries argued in these
papers has been shown not to hold for specific bath-coupled constructions
\cite{Geng:2025gns}.}.

This provides an interesting perspective on our result. If the conjecture
is true, then a UV completion preserving the
relevant radiative commutativity must be non-holographic.  An example would be the Fock
quantization of the boundary $f^+,f^-$ modes. Indeed, such a quantization does not by itself
provide a microscopic accounting of the Bekenstein--Hawking entropy.

Interestingly, the HoI results established here can hold
in theories that are non-holographic in the above sense. This shows a
distinction between the notion of holography that appears in HoI and
the stronger notion which requires a microscopic account of finite
black hole entropy.

The conjecture would extend the conclusion beyond these examples to
any UV completion preserving the relevant commutativity in this setting.

\section{Conclusion}
\label{sec:conclusion}

In this paper, we have studied the asymptotic algebras and Holography of Information for the CGHS model coupled to a massless scalar field. The CGHS model provided a setting where the algebraic structures simplify---there is no independent propagating gravitational data, no soft vacuum degeneracy, and the two chiral radiative algebras can be described explicitly. Our analysis proceeded in two stages: first establishing HoI for the CGHS model, and then examining the tension between the HoI and island approaches. We treated both the complete two-sided boundary and the restriction to the right boundary.

The classical analysis began with a covariant phase space construction for the CGHS model at the right future null infinity $\mathcal{I}^+_R$. In the linear-dilaton-vacuum collapse sector with fixed homogeneous dilaton data, the key structural feature is that the dilaton lies along the null directions of the presymplectic form, so the reduced phase space is parametrized by the matter field alone. We then constructed the radiative phase space at $\mathcal{I}^+_R$, which we found was built entirely from $f^-$ and its derivatives. We obtained the boundary Hamiltonian, the right-cut charge, the complete-cut Bondi mass, and the algebra of asymptotic symmetries. We noted that, unlike in four dimensions, the asymptotic symmetries act homogeneously on the fundamental field, so the charges are quadratic in $\partial_- f^-$ and no soft vacuum degeneracy arises.

An important qualification is that, unlike for 4-dimensional asymptotically flat spacetimes, there is no unique gravitational completion of the radiative algebra in the CGHS model. We have chosen to adjoin the cut charge $\QB$ on a one-sided boundary and the complete-cut Bondi mass $M_{\rm tot}$ in the two-sided description. These quantities are cut-local, can be read directly from the asymptotic dilaton, obey the Bondi balance law, and generate the relevant null translations of the radiative observables. They are therefore the closest analogues of the asymptotic Hamiltonian used in the standard HoI construction. Our HoI results should consequently be understood as statements about this gravitational completion. Other choices of boundary observables would define different algebras and need not lead to the same conclusions.

Under assumptions analogous to those made in four-dimensional asymptotically flat gravity, we proved that the full state can be recovered from arbitrarily small neighbourhoods of the two past corners. The relevant gravitational observable is the complete-cut Bondi mass $M_{\mathrm{tot}}(u,v)$. We then restricted the construction to $\mathcal I^+_R$. The one-sided completed algebra is
$B(\Hfm)\,\bar\otimes\,\mathcal C_L$, where
$\mathcal C_L=\{H_L\}''$. It recovers the right-moving state and the
spectral data of $H_L$, but not the full left-moving state. The entropy of
any right past segment, once a fixed entropy convention is chosen, is
independent of its upper endpoint. The corresponding Page curve is
therefore flat.

 We further argued that these results extend to situations where a horizon is present but no singularity reaches $\mathcal{I}^+_R$. The quantum state on the complete physical $\mathcal{I}^+_R$ can be recovered from an arbitrarily small neighbourhood of its past boundary---but this state is mixed, since the Hawking partners behind the horizon have been traced over. The Page curve remains flat but at a nonzero value of the entropy. This provides a clean example of a setting where one can retrieve the quantum state from a small boundary neighbourhood, yet the trans-horizon information remains lost to an observer confined to the physical boundary. 

The second part of our analysis examined the tension between HoI and the island formula. The island computation for an evaporating black hole in the closely related RST model, carried out in \cite{Hartman:2020swn}, produces a Page curve that rises and then falls, with the late-time decrease driven by an island region extending into the black hole interior. Both \cite{Antonini:2025sur} and \cite{Geng:2026asi} agree that the island entropy should be identified with the von Neumann entropy of the radiative algebra $\mathcal{A}_{\text{rad}}$. Our results establish that in CGHS, $\mathcal{A}_{\text{rad}}(\mathcal{I}^+_R)$ is built entirely from $f^-$ modes and their derivatives. The gravitationally completed algebra $\mathcal{A}_{\text{grav}}(\mathcal{I}^+_R)$ carries left-sector information only through $\mathcal C_L$ and does not determine the full $f^+$ quantum state. Yet the island region contains both $f^+$ and $f^-$ modes, so the island entropy accounts for information that can not be accessed by $\mathcal{A}_{\rm rad}(\mathcal{I}^+_L)$ provided it commutes with $\mathcal{A}_{\rm rad}(\mathcal{I}^+_R)$.

The inference, assuming that the island formula can be shown to hold rigorously for CGHS model, is that the Euclidean path integral that leads to the island formula defines a quantization that does not respect the commutativity of $\mathcal{A}_{\rm rad}(\mathcal{I}^+_R)$ and $\mathcal{A}_{\rm rad}(\mathcal{I}^+_L)$. This poses an interesting puzzle about how exactly the quantization violates the commutativity of the two algebras which requires future investigation.

It is natural to ask what lessons the two-dimensional analysis offers for the physically relevant case of four-dimensional asymptotically flat gravity.  Several features are likely to carry over. The observation that HoI holds even in the presence of horizons, yielding a flat but nonzero Page curve for the mixed state on a complete future null infinity, is a structural feature that would be interesting to check in four dimensions. 

Second, CGHS model provides an example where Holography of Information holds in a theory that is non-holographic in the sense of \cite{Harlow:2020bee, Harlow:2021lpu}. This suggests that the notion of holography at work in HoI is logically independent of the notion that underpins the island formula.  If the CGHS lesson carries over, HoI could be a generic feature of gravitational theories with suitable asymptotic structure, while the island formula and the rising-and-falling Page curve would be specific to theories with a holographic UV completion. It will be interesting to investigate this possibility in the future.

 \begin{acknowledgments}
We would like to thank Alok Laddha, Henry Maxfield, Ronak Soni and Madhavan Varadarajan for helpful discussions. AI has been used for language-editing.  
 \end{acknowledgments}

\bibliography{main}
\appendix

\section{Variation of the CGHS action}
\label{appendixA}

We write the CGHS action as
\begin{equation}
S=\frac12\int d^2x\,\sqrt{-g}\left[
e^{-2\phi}\left(R+4(\nabla\phi)^2+4\lambda^2\right)
-\frac12(\nabla f)^2
\right].
\end{equation}
We define
\begin{equation}
\Phi=e^{-2\phi},\qquad
T_{\mu\nu}^{(f)}
=\nabla_\mu f\nabla_\nu f
-\frac12 g_{\mu\nu}(\nabla f)^2 .
\end{equation}
With the normalization of the matter action above, the tensor used in the
main-text field equations is
$T_{\mu\nu}=\frac12T_{\mu\nu}^{(f)}$.
Let $\mathbf L$ denote the Lagrangian density. We vary the inverse metric
$g^{\mu\nu}$, together with $\phi$ and $f$, and write
\begin{equation}
\delta \mathbf L
=
\sqrt{-g}\left(
E_{\mu\nu}\delta g^{\mu\nu}
+E_\phi\delta\phi
+E_f\delta f
\right)
+\partial_\mu\left(\sqrt{-g}\,\Theta^\mu\right).
\end{equation}
The three terms in the Lagrangian density are
\begin{equation}
\mathbf L_R=\frac12\sqrt{-g}\,\Phi R,\qquad
\mathbf L_\phi
=2\sqrt{-g}\,\Phi\left((\nabla\phi)^2+\lambda^2\right),
\qquad
\mathbf L_m=-\frac14\sqrt{-g}\,(\nabla f)^2.
\end{equation}

The variation of the curvature term is
\begin{align}
\delta\mathbf L_R
&=
\sqrt{-g}\left[
\frac12\Phi G_{\mu\nu}
+\frac12\left(
g_{\mu\nu}\nabla^2\Phi-\nabla_\mu\nabla_\nu\Phi
\right)
\right]\delta g^{\mu\nu}
-\sqrt{-g}\,\Phi R\,\delta\phi
\nonumber\\
&\quad
+\partial_\mu\left(\sqrt{-g}\,\Theta_R^\mu\right),
\\
\Theta_R^\mu
&=
\frac12\left[
\Phi\nabla^\mu(g_{\alpha\beta}\delta g^{\alpha\beta})
-\Phi\nabla_\nu\delta g^{\mu\nu}
-\nabla^\mu\Phi\,g_{\alpha\beta}\delta g^{\alpha\beta}
+\nabla_\nu\Phi\,\delta g^{\mu\nu}
\right].
\end{align}
In two dimensions $G_{\mu\nu}=0$.

The variation of the dilaton kinetic and potential terms is
\begin{align}
\delta\mathbf L_\phi
&=
\sqrt{-g}\left[
2\Phi\nabla_\mu\phi\nabla_\nu\phi
-\Phi g_{\mu\nu}\left((\nabla\phi)^2+\lambda^2\right)
\right]\delta g^{\mu\nu}
\nonumber\\
&\quad
+4\sqrt{-g}\,\Phi
\left[(\nabla\phi)^2-\nabla^2\phi-\lambda^2\right]\delta\phi
+\partial_\mu\left(\sqrt{-g}\,\Theta_\phi^\mu\right),
\\
\Theta_\phi^\mu
&=
4\Phi\nabla^\mu\phi\,\delta\phi
=\nabla^\mu\Phi\,\frac{\delta\Phi}{\Phi}.
\end{align}

The variation of the matter term is
\begin{align}
\delta \mathbf L_m
&=
-\frac14\sqrt{-g}\,T_{\mu\nu}^{(f)}\delta g^{\mu\nu}
+\frac12\sqrt{-g}\,\nabla^2 f\,\delta f
+\partial_\mu\left(\sqrt{-g}\,\Theta_m^\mu\right),
\\
\Theta_m^\mu
&=
-\frac12\nabla^\mu f\,\delta f .
\end{align}

Adding the three pieces gives
\begin{align}
E_\phi
&=
-\Phi\left[
R+4\nabla^2\phi-4(\nabla\phi)^2+4\lambda^2
\right],
\\
E_f
&=
\frac12\nabla^2 f,
\\
E_{\mu\nu}
&=
\Phi\left[
\nabla_\mu\nabla_\nu\phi
-g_{\mu\nu}\nabla^2\phi
+g_{\mu\nu}\big((\nabla\phi)^2-\lambda^2\big)
\right]
-\frac14 T_{\mu\nu}^{(f)} .
\end{align}
Thus the equations of motion are
\begin{align}
R+4\nabla^2\phi-4(\nabla\phi)^2+4\lambda^2&=0,
\\
\nabla^2 f&=0,
\\
2e^{-2\phi}
\left[
\nabla_\mu\nabla_\nu\phi
-g_{\mu\nu}\nabla^2\phi
+g_{\mu\nu}\big((\nabla\phi)^2-\lambda^2\big)
\right]
-\frac12T_{\mu\nu}^{(f)}
&=0 .
\end{align}
The last equation is the metric equation with the matter contribution included.

Finally, the full presymplectic potential is
\begin{equation}
\Theta^\mu
=
\frac12\left[
\Phi\nabla^\mu(g_{\alpha\beta}\delta g^{\alpha\beta})
-\Phi\nabla_\nu\delta g^{\mu\nu}
-\nabla^\mu\Phi\,g_{\alpha\beta}\delta g^{\alpha\beta}
+\nabla_\nu\Phi\,\delta g^{\mu\nu}
+2\nabla^\mu\Phi\,\frac{\delta\Phi}{\Phi}
-\nabla^\mu f\,\delta f
\right].
\end{equation}

This is \eqref{presymplectic} in the main text. 

\section{Algebraic Proof of Holography of Information}
\label{app:algebraic}

In this appendix we give an alternative proof of One-sided Results~1
and~3. The proof does not use the tensor product decomposition
$\mathcal H=\Hfp\otimes\Hfm$. We use only the commutativity of
$\mathcal C_L$ with the right radiative algebra,
Assumption~\ref{ass:right-cut-relations}, and the renormalized operator
form of \eqref{eq:QB-flux}.

All unbounded charges below are understood through their spectral
projections. For a finite cut $q$, we assume a renormalized self-adjoint
operator $M_R(q)$ affiliated with $\Arad(\scri^+_R)$. We denote it by
\begin{equation}
M_R(q)=\int_q^\infty dx^-\,T_{--}(x^-)
\label{eq:algebraic-MR}
\end{equation}
with the regulated convention stated in Section~\ref{subsec:right-cut-charge}.
We assume that the operators have a common invariant core and that $H_L$
and $M_R(q)$ strongly commute. The
operator $\QB(q)$ is the self-adjoint closure of their sum. The finite-cut
relation is
\begin{equation}
H_L=\QB(q)-M_R(q).
\label{eq:algebraic-HL}
\end{equation}

We first recover the global algebra. Every
$\QB(q)=H_L+M_R(q)$ is affiliated to
$\Arad(\scri^+_R)\vee\mathcal C_L$. This gives one inclusion.
Conversely, $\QB(q)$ and $M_R(q)$ are affiliated to
$\Agrav(\scri^+_R)$. Strong commutativity gives
\begin{equation}
e^{itH_L}=e^{it\QB(q)}e^{-itM_R(q)}
\in\Agrav(\scri^+_R).
\label{eq:algebraic-HL-unitary}
\end{equation}
It follows that $\mathcal C_L\subset\Agrav(\scri^+_R)$. Hence
\begin{equation}
\Agrav(\scri^{+}_{R})
=\Arad(\scri^{+}_{R})\vee\mathcal C_L.
\label{eq:algebraic-global}
\end{equation}

\textbf{Result 1$'$.} For every $\epsilon>0$,
\begin{equation}
\Agrav(I_{\epsilon})=\Agrav(\scri^{+}_{R}).
\label{eq:algebraic-result}
\end{equation}

\begin{proof}
Isotony gives
$\Agrav(I_\epsilon)\subset\Agrav(\scri^+_R)$. For the reverse
inclusion, choose $u\in I_\epsilon$. The unitary generated by $\QB(u)$
translates smeared right radiative fields supported to the future of $u$.
These unitaries and additivity give
\begin{equation}
\Arad(\scri^+_R)\subset\Agrav(I_\epsilon).
\end{equation}
Choose a finite $q\in I_\epsilon$. Then $\QB(q)$ is affiliated to
$\Agrav(I_\epsilon)$. The operator $M_R(q)$ is affiliated to the same
algebra because it is affiliated to $\Arad(\scri^+_R)$. Equation
\eqref{eq:algebraic-HL-unitary} now gives
$\mathcal C_L\subset\Agrav(I_\epsilon)$. Equation
\eqref{eq:algebraic-global} proves the reverse inclusion.
\end{proof}

\textbf{Result 3$'$.} For any normal state $\omega$, the restricted state
$\omega|_{\Agrav(I_{<u})}$ is independent of $u$.

\begin{proof}
For any $u$, choose $\epsilon$ so that $I_{\epsilon}\subset I_{<u}$.
By isotony and Result~1$'$,
\begin{equation}
\Agrav(\scri^{+}_{R})
=\Agrav(I_{\epsilon})
\subseteq\Agrav(I_{<u})
\subseteq\Agrav(\scri^{+}_{R}).
\label{eq:algebraic-past}
\end{equation}
Thus $\Agrav(I_{<u})=\Agrav(\scri^{+}_{R})$ for every $u$. The restricted
functional is the same state on the same algebra.
\end{proof}

The content of Result~3$'$ is equality of the restricted states. If an
entropy is assigned using a fixed entropy convention, it is therefore
independent of $u$. The algebra-equality statements
underlying One-sided Results~2 and~4 are also algebraic. Thus, under the
assumptions above, the one-sided HoI result does not require Hilbert space
factorization.

\end{document}